\newcommand{\techreport}[1]{#1}
\newcommand{\paper}[1]{}
\techreport{
\documentclass[times, 12pt]{article} 
\usepackage{a4wide}
}
\paper{
\documentclass[times, 10pt,twocolumn]{article} 
\usepackage{latex8}
}

\usepackage{times}
\usepackage{epsfig}
\usepackage{latexsym}
\usepackage{txfonts}
\usepackage{color}
\usepackage{amssymb}
\usepackage{defpaper}

\paper{
\pagestyle{empty}
}

\techreport{
\newcommand{\Section}[1]{\section{#1}}
\newcommand{\SubSection}[1]{\subsection{#1}}
}

\long\def\symbolfootnote[#1]#2{\begingroup%
\def\thefootnote{\fnsymbol{footnote}}\footnote[#1]{#2}\endgroup}

\newcommand{\refsec}[1]{Section~\ref{#1}}

\newcommand{\reffig}[1]{Figure~\ref{#1}}
\newcommand{\refexample}[1]{Example~\ref{#1}}
\newcommand{\refproblem}[1]{Problem~\ref{#1}}

\newcommand{\algsize}[1]{\footnotesize{#1}}

\newcommand{\vspc}[1]{\vspace{#1}}

\techreport{
\newcommand{\fullpaper}[1]{#1}
\newcommand{\shortpaper}[1]{}
}
\paper{
\newcommand{\fullpaper}[1]{}
\newcommand{\shortpaper}[1]{#1}
}

\begin{document}

\paper{
\title{Reducing Order Enforcement Cost in Complex Query Plans}

\author{
%Ravindra Guravannavar \hspace{5ex} S Sudarshan \hspace{5ex} Ch. Sobhan Babu\\
\hspace{-5ex}Ravindra Guravannavar \hspace{8ex} S Sudarshan \\
\hspace{-2ex}Indian Institute of Technology Bombay\\
%\hspace{-2ex}\{ravig,sudarsha,sobhan\}@cse.iitb.ac.in
\hspace{-2ex}\{ravig,sudarsha\}@cse.iitb.ac.in
}
\maketitle
\thispagestyle{empty}
}
\techreport{
\begin{titlepage}
\begin{center}
\vspace{2in}
{\Large \bf Reducing Order Enforcement Cost in Complex Query Plans}\\
\vspace{1in}  
{\Large Technical Report} \\
\vspace{0.6in}
By\\ 
\vspace{0.3in}
{\large Ravindra N. Guravannavar} \\
\vspace{5pt}  
{\large S. Sudarshan}\\
\vspace{5pt}  
{\large Ajit A. Diwan}\\
\vspace{5pt}  
{\large Ch. Sobhan Babu}\\
%\vspace{0.05in}
%{\bf Roll No : 03405702} \\
%\vspace{0.50in}{\bf Advisor} \\
%\vspace{0.1in}{\bf Prof. S. Sudarshan}\\
\vspace{1in}
%\parbox{\logowidth}{\iitbseal}\\[3mm]
Department of Computer Science and Engineering \\
Indian Institute of Technology Bombay \\
Mumbai \\
%August 2006 \\
%\today \\
\end{center}
\end{titlepage}

\pagenumbering{roman}
%\input{abstract.tex}
%\setcounter{page}{0}
%\newpage

\newpage
%\tableofcontents
\pagenumbering{arabic}
}

\paper{
\begin{abstract}
Algorithms that exploit sort orders are widely used to
implement joins, grouping, duplicate elimination and other set operations.
Query optimizers traditionally deal with sort orders by using 
the notion of interesting orders.
The number of interesting orders is unfortunately factorial in the number of 
participating attributes.
\eat{
Operators such as merge based join, union and sort based 
grouping, duplicate elimination are agnostic to the exact sort order
as long as the sort orders of the inputs match. 
Optimization techniques proposed earlier exploit this fact for single 
input operations, but are not applicable to the general case.  
}
Optimizer implementations use heuristics to prune 
the number of interesting orders, but the quality of the heuristics is unclear.
Increasingly complex decision support queries
and increasing use of covering indices, which provide multiple alternative
sort orders for relations, motivate us to better address the problem
of optimization with interesting orders.

We show that even a simplified version of optimization
with sort orders is NP-hard and provide principled heuristics
for choosing interesting orders. 
\eat{
We  take into account issues such as (i) added choices of sort orders 
for base relations due to the use of query covering indices 
(ii) sort orders that partially match an order requirement 
(iii) requirement of same sort order from multiple inputs 
(e.g., merge based join, union) and 
(iv) common attributes between multiple joins, grouping and set operations.
}
We have implemented the proposed techniques in a Volcano-style 
cost-based optimizer, and our performance study shows significant
improvements in estimated cost.  
We also executed our plans on a widely used commercial database system, 
and on PostgreSQL, and found that actual execution times for our
plans were significantly better than for plans generated by those systems
in several cases.
%\noindent
%\hl {TODO: Mention contribution for exploiting partial sort orders in evaluation 
%and optimization}

\end{abstract}
}

\techreport{
\begin{abstract}

\end{abstract}
\newpage
}
\paper{
\Section{Introduction} 
\label{sec:intro}
Decision support queries, extract-transform-load (ETL) 
operations, data cleansing and integration often use complex joins, aggregation,
set operations and duplicate elimination. 
Sorting based query processing algorithms 
for these operations are well known. Sorting based algorithms
are quite attractive when physical sort orders of one or more base relations 
fulfill the sort order requirements of operators either completely or partially.
Further, secondary indices that cover a query\footnote{{\em i.e.,} 
contain all attributes of the relation that are used in the query} 
are being increasingly used in read-mostly environments. 
Query covering indices
make it very efficient to obtain desired sort orders without accessing 
the data pages. These factors make it possible for sort based plans to 
significantly outperform hash based counterparts. 
\eat{Queries in the
afore mentioned applications often have many attributes involved
in the join predicates and group-by clause with several attributes 
being common to multiple joins, grouping and explicit order-by clause. 
Hence, in a carefully chosen plan it is possible for a single sort order
to fulfill order requirements of multiple operators.
}

The notion of interesting orders~\cite{SEL:SIGMOD79} has allowed optimizers to consider
plans that could be locally sub-optimal, but produce orders that are
beneficial for other operators, and thus produce a better plan overall.
However, the number of interesting orders for most operators is factorial
in the number of attributes involved. This is not acceptable as queries 
in the afore mentioned applications do contain large number of attributes 
in joins and set operations. 
\eat{Malkemus et.al.~\cite{MALK:SIGMOD96} 
mention this problem and propose an approach of using a flexible order 
specification to reduce the number of interesting orders for certain
operations, such as grouping, which can work with any sort order on
the input. However, their approach works only for single input operators. 
Sorting based algorithms for binary operators 
such as join and union,
though agnostic to the exact input order, require a matching order from multiple
inputs, which makes the problem harder to solve. 
}
%We show that even simplified
%version of the problem of choosing interesting orders is NP-hard and 
%give principled heuristics.

\eat{
However, many operators like sort based join, duplicate elimination, group-by 
and merge based set operations (e.g., union)
%do not have a specific order requirement and 
are agnostic to the exact sort order and are applicable when their 
inputs satisfy any sort order on the relevant attributes. This leads 
to a factorial number of
interesting orders to choose from. 
}
\eat{
The variation in plan cost due to the choice of different interesting 
orders could be very large. A carefully chosen sort order 
that exploits
clustering and covering indices, and additionally exploits commonalities 
between order requirements
of multiple operators, can perform significantly better than a plan
with na\"ively chosen orders. Sorting based algorithms for binary operators 
such as join and union,
though agnostic to the exact input order, require a matching order from multiple
inputs, which makes the problem harder to solve.
}

In this paper we consider the problem of optimization taking sort
orders into consideration.  We make the following technical contributions:

\begin{enumerate}
\item Often order requirements of operators are partially satisfied by 
inputs. For instance, consider a merge-join with join predicate
$(r.c_1=s.c_1$ {\em and} $r.c_2=s.c_2)$. A clustering index on $r.c_1$ (or on
$r.c_2$ or $s.c_1$ or $s.c_2$) is helpful in getting the
desired order efficiently; a secondary index that covers the query has
the same effect. 
%Partial sort orders, when exploited 
%properly, can yield benefits by avoiding run-generation I/O 
%of external sort. 

We highlight (in \refsec{sec:pso}) the need for exploiting partial sort orders and 
show how a minor modification to the standard
replacement selection algorithm can avoid run generation I/O completely
when input is known to have a partial sort order. Further, we
extend a cost-based optimizer to take into account partial sort
orders.

\item We consider 
%(in Sections~\ref{sec:nphard} and \ref{sec:ford}) 
operators with flexible order requirements and 
address the problem of choosing good
interesting orders so that complete or partial sort orders already available
from inputs can be exploited. 
\begin{itemize}
    \item In Sections~\ref{sec:nphard} we show that a special case of
    finding optimal sort orders is NP-hard and give a 2-approximation
    algorithm to choose interesting sort orders for a join tree.
    \item In \refsec{sec:ford} we address a more general case of the 
    problem. In many cases, the knowledge of indices and available 
    physical operators in the system allows us to narrow down the 
    search space to a small set of orders.  We formalize this idea (in 
    \refsec{subsec:favord}) through 
    the notion of {\em favorable orders}, and propose a heuristic to 
    efficiently enumerate a small set of promising sort orders.
    Unlike heuristics used in optimizer implementations,
    our approach takes into account issues such as $(i)$ added choices of sort orders 
    for base relations due to the use of query covering indices 
    $(ii)$ sort orders that partially match an order requirement 
    $(iii)$ requirement of same sort order from multiple inputs 
    ({\em e.g.,} merge based join, union) and 
    $(iv)$ common attributes between multiple joins, grouping and set operations.
    
    In \refsec{subsec:optext} we also show how to integrate our extensions
    into a cost-based optimizer.
\end{itemize}

\item We present experimental results (in \refsec{sec:perf})
evaluating the benefits of the proposed
techniques. We compare the plans generated by our optimizer with those of 
three widely used database systems and show significant benefits
due to each of our optimizations.
\end{enumerate}

\eat{
The rest of this paper is organized as follows. In \refsec{sec:relwork}, we 
describe and compare related work. 
%\refsec{sec:sysmodel} describes our system model and notation used throughout the paper. 
In \refsec{sec:pso} 
we describe how partial sort orders can be exploited during sorting
and changes to a cost-based optimizer to account for their benefits 
during plan generation. In \refsec{sec:nphard}, we
show that a special case of the problem of selecting globally optimal 
sort orders is NP-hard and give a 2-approximation algorithm 
to handle the case. Although the problem is intractable, the knowledge of available indices and properties of physical
operators allow us to provide a good heuristic approach that we 
describe in \refsec{sec:ford}. We present our experimental results in
\refsec{sec:perf} and conclude in \refsec{sec:concl}.
}

\eat{
The issue of choosing interesting orders from a factorial size space
was briefly mentioned by Malkemus et.al.~\cite{MALK:SIGMOD96}. They 
propose an approach of using a flexible order specification. Such an
approach is not applicable in general and works only for single input
operators. More on related work can be found in \refsec{sec:relwork}
}

}
\techreport{

}
%\input{section3.tex}
% comparison with adaptive sorting algorithms that do good when the number of 
% inversions is less
\Section{Related Work} 
\label{sec:relwork}
\eat{
External sorting algorithms have been studied extensively but in isolation. 
The standard replacement selection~\cite{KNUTH} for run formation well
adapts with the extent to which input is presorted. In the extreme case, 
when the input is 
fully sorted, it generates a single run on the disk and avoids merging
altogether. Per-{\AA}ke Larson~\cite{PAUL} revisits run formation in the
context of query processing and extends the standard replacement selection
to handle variable length keys and to improve locality of reference (reduced
cache misses). Estivill-Castro and Wood~\cite{WOOD} provide a survey of
adaptive sorting algorithms. The technique we propose to exploit partial
sort orders is a specific optimization in the context of multi-key external
sorting. We observe that, by exploiting prior knowledge of partial sort
order of input, it is possible to eliminate disk I/O altogether and have
a completely pipelined execution of the sort operator. A significant 
benefit of this optimization is that the sort operator can start producing
the tuples very early, which is important for Top-K queries and queries
for which the user retrieves only few answer tuples.
}

Both System R~\cite{SEL:SIGMOD79} and Volcano~\cite{GRA:ICDE93} optimizers
consider plans
that could be locally sub-optimal but provide a sort order of interest to other
operators and thus yield a better plan overall. However, the papers assume operators
have one or few {\em exact} sort orders of interest. This is not true of 
operators like merge-join, merge-union, grouping and duplicate elimination that have
a factorial number of interesting orders. Heuristics such as the PostgreSQL
heuristic we shall see, are commonly used by optimizers. Details of the heuristics
are publicly available only for PostgreSQL. Further, System R and Volcano
optimizers consider only those sort orders as useful that completely meet 
an order requirement. Plans that partially satisfy a sort order requirement
are not handled. 
\paper{In this paper we address these two issues.}
\techreport{In this report we address these two issues.}

The seminal work by Simmen et.al.~\cite{MALK:SIGMOD96} describes techniques 
to infer orders from functional dependencies and predicates applied and
thereby avoids redundant sort enforcers in the plan. 
The paper briefly mentions the problem of non-exact sort order
requirements and mentions an approach of 
%pushing down a flexible order
%requirement. The approach, instead of generating a concrete interesting
%order, propagates 
propagating an order specification that allows any permutation
on the attributes involved.
Though such an approach is possible for single input 
operators like group-by, it cannot be used 
for operators such
as merge-join and merge-union for which the order guaranteed by both inputs must 
match. Moreover, the paper does not make it clear how the 
flexible order requirements are combined at other joins and group-by 
operators. 
\eat{As we show in this paper, choosing optimal sort 
orders for a set of merge-joins is NP-Hard and requires careful 
attention. 
} 
Simmen et.al.~\cite{MALK:SIGMOD96} mention that  the approach of 
carrying a flexible order
specification also increases the complexity of the code significantly.
Our techniques do not use flexible order specifications and hence can be 
incorporated into an existing optimizer with minimal changes. 
Further, our techniques work uniformly across all types of operators
that have a flexible order requirement.
Work on inferring orders and groupings~\cite{MALK:SIGMOD96}~\cite{MITCH:VLDB03}
~\cite{MOER:VLDB04}~\cite{MOER:ICDE04}
is independent and complementary to our work.

%TODO: terms to introduce: partial sorting, prefix segment size
\Section{Exploiting Partial Sort Orders} 
\label{sec:pso}
Often, sort order requirements of operators are partially satisfied
by indices or other operators in the input subexpressions. A prior
knowledge of partial sort orders available from inputs allows us 
to efficiently produce the required (complete) sort order more
efficiently. When operators have flexible order requirements,
it is thus important to choose a sort order that makes maximum
use of partial sort orders already available. We motivate the
problem with an example.
Consider the query shown in ~\refexample{example1}. Such queries frequently 
arise in consolidating data from multiple sources. The join predicate 
between the two {\em catalog} tables involves four attributes and two of
these attributes are also involved in another join with the {\em rating}
table. 
Further, the order-by clause asks for sorting on a large number of columns 
including the columns involved in the join predicate. 

\paper{
\newtheorem{eioexample}{Example}
\begin{eioexample}
\label{example1}
\noindent A query with complex join condition
%{\em 
{\footnotesize
\begin{tabbing}
xxxxxxxxx\=xxxxxxxxx\=xxxxxxxxx\=\kill
SELECT c1.make, c1.year, c1.city, c1.color, c1.sellreason, \\
xxxx\=xxxxxxxxx\=xxxxxxxxx\=\kill
       \> c2.breakdowns, r.rating \\
xxxxxxxxx\=xxxxxxxxx\=xxxxxxxxx\=\kill
FROM  catalog1 c1, catalog2 c2, rating r \\
xxxxxxxxx\=xxxxxxxxx\=xxxxxxxxx\=\kill
WHERE c1.city=c2.city AND c1.make=c2.make AND c1.year=c2.year \\
xxxx\=xxxxxxxxx\=xxxxxxxxx\=\kill
       \> AND c1.color=c2.color AND c1.make=r.make and c1.year=r.year  \\
xxxxxxxxx\=xxxxxxxxx\=xxxxxxxxx\=\kill
ORDER BY c1.make, c1.year, c1.color, c1.city, \\
xxxx\=xxxxxxxxx\=xxxxxxxxx\=\kill
       \> c1.sellreason, c2.breakdowns, r.rating;
\end{tabbing}
}
%}
\end{eioexample}
}

\techreport{
\newtheorem{eioexample}{Example}
\begin{eioexample}
\label{example1}
\noindent A query with complex join condition
%{\em 
{\footnotesize
\begin{tabbing}
xxxxxxxxx\=xxxxxxxxx\=xxxxxxxxx\=\kill
SELECT \> c1.make, c1.year, c1.city, c1.color, c1.sellreason, 
       c2.breakdowns, r.rating \\
FROM  \> catalog1 c1, catalog2 c2, rating r \\
WHERE \> c1.city=c2.city AND c1.make=c2.make AND c1.year=c2.year
       AND c1.color=c2.color AND \\
   \> c1.make=r.make and c1.year=r.year  \\
ORDER BY c1.make, c1.year, c1.color, c1.city,
         c1.sellreason, c2.breakdowns, r.rating;
\end{tabbing}
}
%}
\end{eioexample}
}

\paper{
\begin{figure} [t]
\centerline{\psfig{file=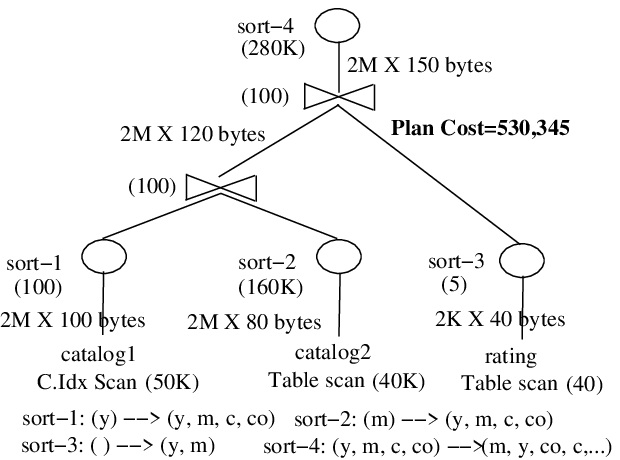}}
\caption{A na\"ive plan for Example \ref{example1}}
\label{fig:naiveplan}
\end{figure}

\begin{figure} [t]
\centerline{\psfig{file=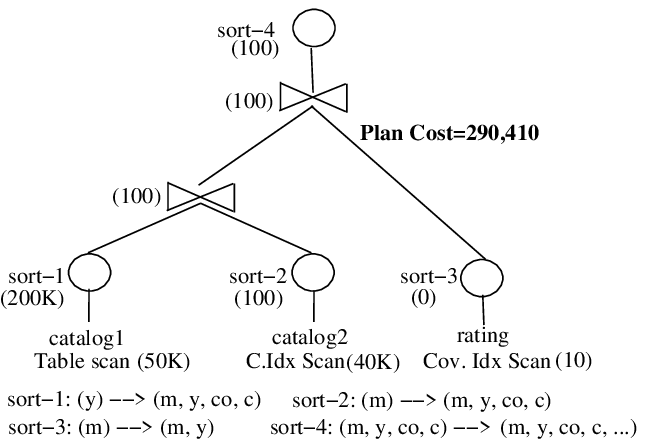}}
\caption{Optimal merge-join plan for Example \ref{example1}}
\label{fig:optplan}
\end{figure}
}

\techreport{
\begin{figure}[ht]
\begin{tabular}{@{}c@{}c}
\begin{minipage}{0.5\textwidth} %DEFINES THE WIDTH OF EACH PICTURE IN % OF PAGE WIDTH
\centerline{\includegraphics{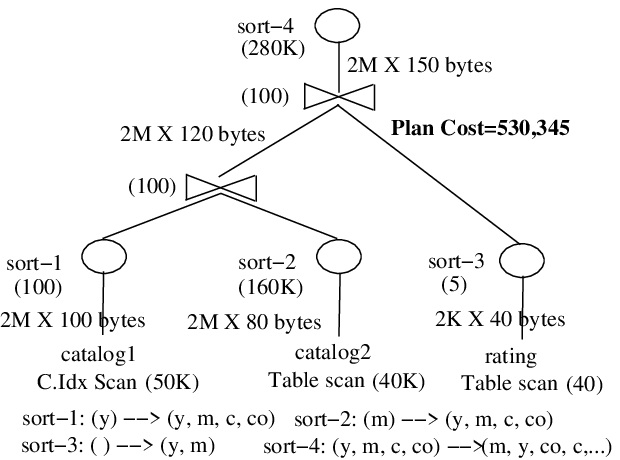}}
\caption{A na\"ive plan}
%for Example \ref{example1}}
\label{fig:naiveplan}
\end{minipage}
&
\begin{minipage}{0.5\textwidth} %DEFINES THE WIDTH OF EACH PICTURE IN % OF PAGE WIDTH
\centerline{\includegraphics{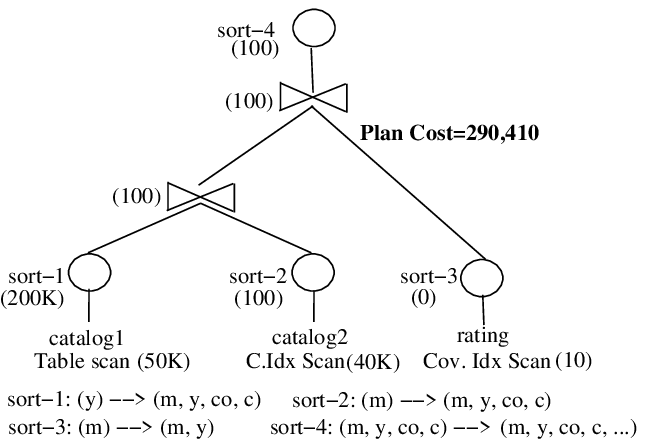}}
\caption{Optimal merge-join plan}
%for Example \ref{example1}}
\label{fig:optplan}
\end{minipage}
\end{tabular}
\end{figure}
}

The two catalog tables contain 2 million records each and have average 
tuple sizes of 100 and 80. We assume a disk block size of 4K bytes and 
10000 blocks (40 MB) of main memory for sorting.  
The table {\em catalog1} is clustered on {\em year} and the
table {\em catalog2} is clustered on {\em make}. The {\em rating} table
has a secondary index on the {\em make} column with the {\em year} and
{\em rating}
columns included in the leaf pages (a covering index).
\reffig{fig:naiveplan} and ~\ref{fig:optplan} show two different plans
for the example query. Numbers in the parentheses indicate estimated 
cost of the operators in number of I/Os (CPU cost is appropriately
translated into I/O cost units). Edges are marked with the number of 
tuples expected
to flow on that edge and their average size.  For brevity, the input
and output orders for the sort enforcers are shown using only the starting 
letters of the column names. Though both 
plans use the same join order and employ sort-merge joins, the
second plan is expected to perform significantly better than the first.

\SubSection{Changes to External Sort}
\label{subsec:extsort}
%The sort enforcer should exploit the partial sort order already known to 
%hold on its input and produce the required (complete) sort order 
%efficiently - in most cases, without writing any temporary runs to the 
%disk. 
External sorting algorithms have been studied extensively but in isolation. 
The standard replacement selection~\cite{KNUTH} for run formation well
adapts with the extent to which input is presorted. In the extreme case, 
when the input is 
fully sorted, it generates a single run on the disk and avoids merging
altogether. %Per-{\AA}ke 
Larson~\cite{PAUL} revisits run formation in the
context of query processing and extends the standard replacement selection
to handle variable length keys and to improve locality of reference (reduced
cache misses). Estivill-Castro and Wood~\cite{WOOD} provide a survey of
adaptive sorting algorithms. The technique we propose in this section 
to exploit partial
sort orders is a specific optimization in the context of multi-key external
sorting. We observe that, by exploiting prior knowledge of partial sort
order of input, it is possible to eliminate disk I/O altogether and have
a completely pipelined execution of the sort operator. 
%A significant 
%benefit of this optimization is that the sort operator can start producing
%the tuples very early, which is important for Top-K queries and queries
%for which the user retrieves only few answer tuples.

\noindent We use the following notations: 
We use $o, o_1, o_2$ {\em etc.} to refer to sort orders. Each 
sort order $o$ is a sequence of attributes/columns $(a_1, a_2, \ldots a_n)$. 
We ignore the  sort direction (ascending/descending) as our techniques 
are applicable independent of the sort direction. 

\begin{itemize}
\addtolength{\itemsep}{-5pt}
\small{
\item $\epsilon$  : Empty (no) sort order
\item {\em attrs}$(o)$ : The set of attributes in sort order $o$
\item $|o|$ : Number of attributes in the sort order $o$
\item $o_1 \leq o_2$  : Order $o_2$ subsumes order $o_1$ ($o_1$ is a prefix of $o_2$)
\item $o_1 < o_2$   : Order $o_1$ is a strict prefix of $o_2$ 
}
\end{itemize}

\eat{
\noindent {\em Notations Used}: 
We use $o, o_1, o_2$ {\em etc.} to refer to sort orders. Each 
sort order $o$ is a sequence of attributes/columns $(a_1, a_2, \ldots a_n)$. 
We ignore the  sort direction (ascending/descending) as our techniques 
are applicable independent of the sort direction. 

{ \footnotesize
\begin{flushleft}
\begin{tabular}{|l|l|} \hline
$\epsilon$          & Empty (no) sort order \\
\hline
{\em attrs}$(o)$    & The set of attributes in sort order $o$ \\
\hline
$|o|$               & Number of attributes in the sort order $o$ \\
\hline
$o_1 \leq o_2$      & Order $o_2$ subsumes order $o_1$ ($o_1$ is a prefix of $o_2$)\\
\hline
$o_1 < o_2$         & Order $o_1$ is a strict prefix of $o_2$ \\
\hline
\end{tabular}
\end{flushleft}
}

\vspace{1ex}
}

%Most database systems use replacement-selection~\cite{KNUTH} or its variants
%for run formation during external sort. Although replacement-selection adapts
%well (produces fewer runs) with the degree of sortedness in the input,
%it does not automatically make use of known partial sort orders. 

Consider
a case where the sort order to produce is $(col_1, col_2)$ and the input 
already has the order $(col_1)$. Standard replacement-selection writes a
single large run to the disk and reads it back again; this breaks the pipeline 
and incurs substantial I/O for large inputs. It is not difficult to see 
how the standard
replacement-selection can be modified to exploit the partial
sort orders. Let $o=(a_1, a_2, \ldots a_n)$ be the desired sort 
order and $o'=(a_1, a_2, \ldots a_k)$, $k < n$ be the partial
sort order known to hold on the input. At any point during sorting
we need to retain only those tuples that have the same value for 
attributes $a_1, a_2, \ldots a_k$. When a tuple with a new value for
these set of attributes is read, all the tuples in the heap (or on disk 
if there are large number of tuples matching a given value of 
$a_1, a_2, \ldots a_k$) can be sent to the next operator in sorted
order. Thus in most cases, partial sort orders allow a completely
pipelined execution of the sort. Exploiting partial sort orders
in this way has several benefits:
\begin{enumerate}
\item Let $o=(a_1, a_2, \ldots a_n)$ be the desired sort order and 
$o'=(a_1, a_2, \ldots a_k)$, $k < n$ be the partial sort order known 
to already hold on the input. We call the set of tuples that have 
the same value for attributes $(a_1, a_2, \ldots a_k)$ as a 
{\em partial sort segment}. If each {\em partial sort segment} fits
in memory (which is quite often the case in practice), the entire 
sort operation can be completed without any disk I/O.
\item Exploiting partial sort orders allows us to output tuples 
early (as soon as a new segment starts). In a pipelined execution
this can have large benefits. Moreover, producing tuples early
has immense benefits for Top-K queries and situations where the
user retrieves only some result tuples.
\item Since sorting of each {\em partial sort segment} is done independently,
the number of comparisons are significantly reduced. Note that we 
empty the heap every time a new segment starts and hence insertions
into heap will be faster. In general, independently sorting $k$ segments
each of size $n/k$ elements, has the complexity $O(n\ log(n/k))$ as
against $O(n\ log(n))$ for sorting all $n$ elements.
Further, while sorting each {\em partial sort segment} comparisons 
need to be done on fewer attributes, $(a_{k+1},\ldots a_n)$ in the
above case.
\end{enumerate}

Our experiments (\refsec{sec:perf}) confirm that the benefits of exploiting 
partial sort orders can be substantial and yet none of the systems we evaluated,
though widely used, exploited the partial sort orders. 
%For commercial
%systems we confirmed our claim by using special nested rank queries to
%partially sort multiple groups of tuples and comparing the time taken 
%with an equivalent sort on the same system. For PostgreSQL (version 8.1.3)
%the source code was available and confirmed our claim.

%Apart from having a sort enforcer that exploits partial sort orders, 
%the optimizer needs to be extended to consider the benefit of plans
%that partially guarantee some interesting order.

\SubSection{Optimizer Extensions for Partial Sort Orders}
% to Handle Partial Sort Orders}
\label{subsec:optchangepso}

In this section we assume order requirements of operators are concrete and
only focus on incorporating partial sort orders. We deal with flexible order
requirements in subsequent sections.

\noindent We use the following notations: 
\begin{itemize}
\addtolength{\itemsep}{-5pt}
\small{
\item $o_1 \wedge o_2$  : Longest common prefix between $o_1$ and $o_2$
\item $o_1 + o_2$  : Order obtained by concatenating $o_1$ and $o_2$
\item $o_1 -  o_2$ : Order $o'$ such that $o_2 + o' = o_1$ 
                     (defined only when $o_2 \leq o_1$)
\item {\em coe}$(e, o_1, o_2)$  : The cost of enforcing order $o_2$ on the result 
                     of expression $e$ which already has order $o_1$ 
\item $N(e)$       : Expected size, in number of tuples, of the 
                     result of expression $e$
\item $B(e)$       : Expected size, in number of blocks, of the
                     result of expression $e$
\item $D(e, s)$    : Number of distinct values for attribute(s)
                     $s$ of expression $e$ $(\ =N(\Pi_{s}(e))\ )$
\item $cpu\_cost(e, o)$ : CPU cost of sorting result of $e$ to get order $o$
\item $M$          : Number of memory blocks available for sorting
}
\end{itemize}

\eat{
\noindent {\em Notations Used}:
{ \footnotesize
\begin{flushleft}
\begin{tabular}{|l|l|} \hline
$o_1 \wedge o_2$    & Longest common prefix between $o_1$ and $o_2$ \\
\hline
$o_1 + o_2$         & Order obtained by concatenating $o_1$ and $o_2$\\
\hline
$o_1 -  o_2$        & Order $o'$ such that $o_2 + o' = o_1$ (defined only \\
                    & when $o_2 \leq o_1$) \\
\hline
{\em coe}$(e, o_1, o_2)$  & The cost of enforcing order $o_2$ on the result \\
                          & of expression $e$ which already has order $o_1$ \\
\hline
$N(e)$              & Expected size, in number of tuples, of the \\
                    & result of expression $e$ \\
\hline
$B(e)$              & Expected size, in number of blocks, of the \\
                    & result of expression $e$ \\
\hline
%\end{tabular}
%\end{flushleft}
%}

%{ \footnotesize
%\begin{flushleft}
%\begin{tabular}{|l|l|} \hline

%\hline
$D(e, s)$  & Number of distinct values for attribute(s) \\
           & $s$ of expression $e$ $(\ =N(\Pi_{s}(e))\ )$ \\
\hline
$cpu\_cost(e, o)$ & CPU cost of sorting result of $e$ to obtain \\
                  & order $o$ \\
\hline
\end{tabular}
\end{flushleft}
}
}

The {\em Volcano} optimizer framework~\cite{GRA:ICDE93} assumes an algorithm 
(physical operator) either guarantees a required sort order fully or it 
does not. Further, a physical property enforcer (such as sort) only knows 
the property to be enforced and has no information about the properties 
that hold on its input. The optimizer's cost estimate for the enforcer 
thus depends only on the required output property (sort order).
In order to remedy these deficiencies we extended the optimizer in the 
following way: 
Consider an optimization goal $(e, o)$, where $e$ is the expression 
and $o$ the required output sort order. If the physical operator 
being considered for the logical operator at the root of $e$ guarantees
a sort order $o' < o$, then the optimizer adds a partial sort enforcer 
{\em enf} to enforce $o$ from $o'$.  We use the following cost model
to account for the benefits of partial sorting.

\eat{
{\bf Cost Model} \\
We estimate the cost of sorting the result of expression $e$ with a memory 
of $M$ blocks to obtain an order $o$ as follows:
}

{\small
\[coe(e, \epsilon, o) = \left \{ \begin{array}{ll}
                            \mbox{\em {cpu-cost}}(e, o)  & \mbox{if $B(e)\leq M$} \\
                            B(e)(2\lceil log_{M-1} (B(e)/M)\rceil + 1) & \mbox{otherwise}
                                 \end{array}
                        \right. \]
}

\noindent
If $e$ is known to have the order $o_1$, we estimate the cost of obtaining
an order $o_2$ as follows:

\noindent
$coe(e, o_1, o_2)$ $=\ D(e,\ ${\em attrs}$(o_s)) * coe(e', \epsilon, o_r)$,
where $o_s=o_2 \wedge o_1$, $o_r=o_2-o_s$ and $e'=\sigma_{p}(e)$, where $p$ 
equates attributes in $o_s$ to an arbitrary constant. Intuitively, we consider
the cost of sorting a single {\em partial sort segment} independently and 
multiply it by the number of segments.
Note that we assume 
uniform distribution of values for {\em attrs}$(o_s)$. Therefore, we 
estimate $N(e')=N(e)/D(e,\ ${\em attrs}$(o_s))$ and 
$B(e')=B(e)/D(e,\ ${\em attrs}$(o_s))$.
When the actual distribution of values is available, a more accurate cost model 
that does not rely on the uniform distribution assumption can be used.

\Section{Choosing Sort Orders for a Join Tree}
\label{sec:nphard}

Consider a join expression $e = e_1 \Join e_2$, where $e_1, e_2$ are
input subexpressions and the join predicate is of the form: 
$(e_{1}.a_{1}=e_{2}.a_{1}$ {\em and} $e_{1}.a_{2}=e_{2}.a_{2} 
\ldots $ {\em and} $e_{1}.a_{n}=e_{2}.a_{n})$. Note that, {\em w.l.g.},
we use the same name for attributes being compared from 
either side and we call the set $\{a_1, a_2, \ldots, a_n\}$ 
as the join attribute set. In this case, the merge join 
algorithm has potentially $n!$ interesting sort orders on inputs 
$e_{1}$ and $e_{2}$\hspace*{-1mm}~\footnote{We assume merge-join requires 
sorting on all attributes involved in the join predicate. We do not 
consider orders on subsets of join attributes since the additional cost 
incurred at merge-join matches the benefit of sorting a smaller subset 
of attributes.}. The specific sort order chosen for the merge-join can have 
significant influence on the plan cost due to the following reasons:
$(i)$ Clustering and covering indices, indexed materialized views and
other operators in the subexpressions $e_1, e_2$ can make one sort order 
much cheaper to produce than another. $(ii)$ The merge-join produces 
the same order on its output as the one selected for its inputs. Hence,
a sort order that helps another operator above the merge-join can help
eliminate a sort or just have a partial sort. In this section we show 
that a special case of the the problem of choosing optimal sort orders 
for a tree of merge-joins is {\em NP-Hard} and provide a 2-approximate 
algorithm for the problem. In the next section, we 
describe our heuristics for a more general setting of the problem
in which we make use of the proposed 2-approximate algorithm.

\eat{
In the previous section, when we computed the {\em ford-min} for a join
expression, we made an {\em uninformed} choice of permutation for the
join attributes not involved in an input favorable order. Ideally, we
must choose a permutation that benefits other joins above. We call this
the problem of choosing globally optimal sort orders.
We prove a special case of the problem of choosing globally optimal sort orders for
merge joins to be NP-Hard. We then propose a 2-approximation for 
the problem and incorporate the same during our post-optimization phase.
}

\SubSection{Finding Optimal is NP-Hard}
Consider a join expression $e=R_1 \Join R_2 \Join R_3 \ldots R_n$ and 
a specific join order tree for the expression.  Consider a special
case where all base relations  and intermediate results are of the same 
size and no indices built on the base relations. Now, the problem of 
choosing optimal sort orders for each join requires us to choose permutations
of join attributes such that we maximize the length of longest common prefixes 
of permutations chosen for adjacent nodes. 
\reffig{fig:tree}, shows an example and an optimal solution under the model
where the benefit for an edge is the length of the longest common prefix between the 
permutations chosen for adjacent nodes and we maximize the total benefit. 
The join attribute
set for each join node is shown in curly braces besides the node. Permutations 
chosen in the optimal solution are indicated with angle brackets and the number 
on each edge shows the benefit for that edge. 
%Here benefit is the length of the
%common prefix between the permutations chosen for the adjacent nodes.
Below we state the problem formally.
%and show that it is {\em NP-Hard}.

%This problem can be shown 
%to be {\em NP-Hard} (see \reflemma{lemma1}).
%is equivalent to the problem can be translated to the following problem.

\paper{
\begin{figure} [t]
\centerline{\psfig{file=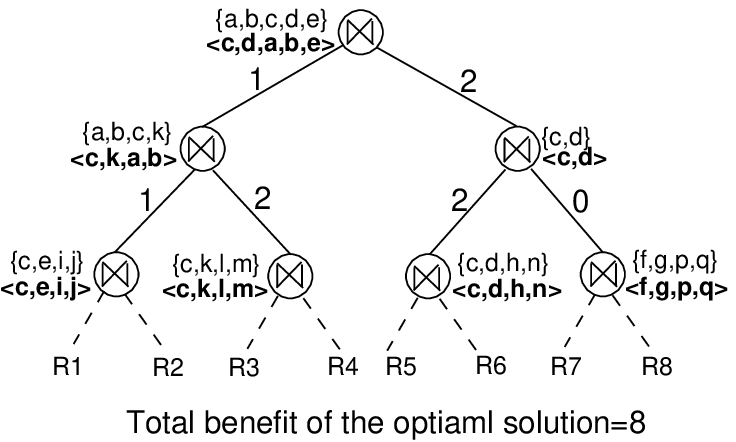}}
%\caption{A special case of choosing globally optimal sort order}
\caption{Optimal sort orders (a special case)}
\label{fig:tree}
\end{figure}
}

\techreport{
\begin{figure} [ht]
%\centerline{\psfig{file=fig/tree.eps}}
\begin{center}
\includegraphics[angle=0,width=4.2in]{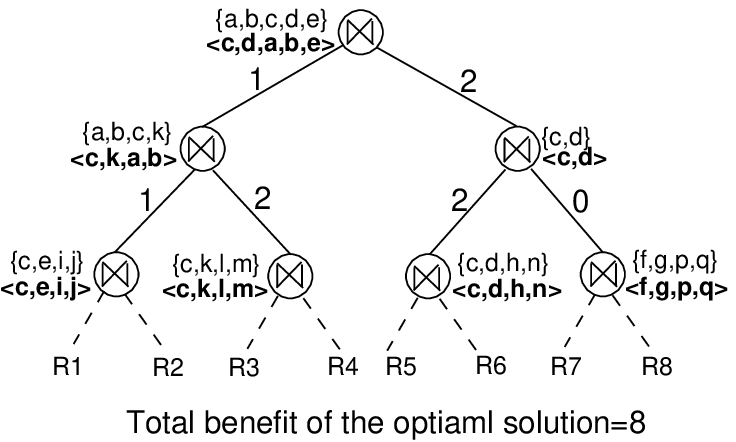}
\end{center}
\caption{A special case of choosing globally optimal sort order}
%\caption{Optimal sort orders (a special case)}
\label{fig:tree}
\end{figure}
}

\noindent
\newtheorem{eioproblem}{Problem}

\eat{
\begin{lemma}
\label{lemma1}
Let $T$ be a binary tree of order $n$, in which each node $v_i$ is associated with an attribute 
set $s_i$. 
%Nodes in the tree correspond to join operations and each node is associated
%a set comprising of the attributes involved in the join predicate.
Finding a sequence of permutations $p_1,p_2 \ldots p_n$, where $p_i$ is a 
permutation of set $s_i$, such that the benefit function $\mathcal{F}$ 
is maximized, is {\em NP-Hard}.

\[ \mathcal{F} = \sum_{\forall v_iv_j\in E(T)} |p_i \wedge p_j| \]

%The benefit of an edge $(v_i, v_j)$ is defined as the length of the maximum 
%common prefix between $p_i$ and $p_j$. 
\end{lemma}
}

\vspace{-2pt}
\begin{eioproblem}
\label{problem1}
Let $T$ be a binary tree of order $n$, with vertex set $V(T)$ and edge set $E(T)$. 
Each node $v_i$ $(i=1,\ldots n)$ is associated with an attribute set $s_i$. 
%Nodes in the tree correspond to join operations and each node is associated
%a set comprising of the attributes involved in the join predicate.
Find a sequence of permutations $p_1,p_2 \ldots p_n$, where $p_i$ is a 
permutation of set $s_i$, such that the benefit function $\mathcal{F}$ 
is maximum.

\[ \mathcal{F} = \sum_{\forall v_iv_j\in E(T)} |p_i \wedge p_j| \]

%The benefit of an edge $(v_i, v_j)$ is defined as the length of the maximum 
%common prefix between $p_i$ and $p_j$. 
\end{eioproblem}

\eat{
\begin{lemma}
\label{lemma1}
The known {\em NP-Hard} SUM-CUT problem~\cite{ACM:LAYOUTSUR} is reducible
to \refproblem{problem1}.
\end{lemma}
}

\begin{theorem}
\label{lemma1}
%The decision problem corresponding to \refproblem{problem1} is NP-hard.
\refproblem{problem1} is NP-hard.
\end{theorem}

\shortpaper{
The known {\em NP-Hard} problem SUM-CUT~\cite{ACM:LAYOUTSUR} is reducible
to \refproblem{problem1}. A formal proof can be found in the full length
paper~\cite{FULLPAPER}.
}

\fullpaper{
\begin{proof}
We give a reduction from the known {\em NP-Hard} problem {\em SUM-CUT}~\cite{ACM:LAYOUTSUR}.

\eat{
\begin{eioproblem}
\label{problem-sumcut}
(SUM-CUT)
{\em
A linear layout or linear arrangement of an undirected graph 
$G=(V, E)$ with $n=|V|$ vertices is a bijective function 
$\varphi : V \rightarrow [n] = {1,\ldots, n}$

Given a layout $\varphi$ of a graph $G=(V,E)$ and an integer $i$, we 
define the set $L(i, \varphi, G)=\{u \in V : \varphi(u) \leq i\}$ and 
the set $R(i, \varphi, G)=\{u \in V : \varphi(u) > i\}$. 
The vertex cut or separation at position $i$ of $\varphi$ is defined as 
\[
\delta(i, \varphi, G)= |\{u \in L(i, \varphi, G) : \exists v \in R(i, \varphi, G) : 
uv \in  E\}|
\]

The SUM-CUT problem calls for finding a linear layout $\varphi$ of $G=(V,E)$ 
that minimizes the function
\[
\sum_{1 \leq i < |V|} \delta(i, \varphi, G)
\]

}
\end{eioproblem}
}
\begin{eioproblem}
\label{problem-sumcut}
(SUM-CUT)
{\em
Given a graph $G=(V, E)$ with $m=|V|$ vertices, number the vertices 
of G as $1,\ldots m$ such that  
$\sum_{1\leq i \leq m} c_i$ is minimized, where $c_i$ is the number of 
vertices numbered $\leq i$ that are connected to at least one vertex
numbered $> i$.
}
\end{eioproblem}

The SUM-CUT problem is equivalent to \refproblem{problem2} given below;
to see the equivalence consider the complement of the graph.

\begin{eioproblem}
\label{problem2}
{\em 
Given a graph $G=(V, E)$ with $m=|V|$ vertices, number the vertices 
of $G$ as $1,2,\ldots ,m$ such that 
$\sum_{1\leq i \leq m} q_i$ is maximized, where $q_i$ is the number 
of vertices that are adjacent to {\em all}
the vertices numbered $1$ to $i$. 
}
\end{eioproblem}

\noindent We now reduce \refproblem{problem2} to \refproblem{problem1}. The 
construction is as follows:

Let the $m$ vertices of the graph be labeled $u_1,\ldots,u_m$. We
construct the binary tree by choosing a vertex set $V(T)$ of size
$2m$, $\{v_1,v_2 \ldots, v_{2m}\}$.
$v_1,\ldots v_m$ are internal vertices and $v_{m+1},\ldots,v_{2m}$
are leaf vertices.
%,u_1,u_2,\ldots,u_n\}$
Let the edge set $E(T)$ be $\{v_iv_{i+1} : 1 \leq i < m\} \bigcup \{v_iv_{m+i}: 1 \leq i \leq m\}$.
%$=\cup_{i<n}v_iv_{i+1} \cup \cup_{i\leq n}v_iv_{n+i}$
Attribute set $s_i$ of internal vertex $v_i$ is $V(G) \cup \mathcal{L}$, 
where $\mathcal{L}$ is an arbitrarily large set disjoint from $V(G)$,
for $1 \leq i \leq m$.
Attribute set $s_i$ of leaf vertex $v_i$ is $\{w: w\in V(G)$ and $wu_{i-m}\in E(G)\}$, 
for $m < i \leq 2m$.
\end{proof}
}

\eat{
We omit a formal proof of \reflemma{lemma1} but give a brief outline below.
%state the sequence of reductions we use.

\noindent
The SUM-CUT problem is  reducible to the following problem:

\begin{eioproblem}
\label{problem2}
Let $G$ be a graph of order $m$, with vertex set $V(G)$ and edge set $E(G)$. 
Let the $m$ vertices be labeled $u_1,\ldots,u_m$. Number the 
vertices of $G$ as $1,2,\ldots ,m$ such that 
$\sum_{1\leq i \leq m} q_i$ is maximized, where $q_i$ is the number of vertices that are 
adjacent to all the vertices numbered $1$ to $i$. 
\end{eioproblem}

Further, \refproblem{problem2} is reducible to \refproblem{problem1}. The 
construction is as follows:

Let the vertex set of tree $T$, $V(T)$, be $\{v_1,v_2 \ldots, v_{2m}\}$,
where $v_1,\ldots v_m$ are internal vertices and $v_{m+1},\ldots,v_{2m}$
are leaf vertices.
%,u_1,u_2,\ldots,u_n\}$
Let the edge set $E(T)$ be $\{v_iv_{i+1} : 1 \leq i < m\} \bigcup \{v_iv_{m+i}: 1 \leq i \leq m\}$.
%$=\cup_{i<n}v_iv_{i+1} \cup \cup_{i\leq n}v_iv_{n+i}$
Attribute set $s_i$ of internal vertex $v_i$ is $V(G) \cup \mathcal{L}$, 
where $\mathcal{L}$ is an arbitrarily large set disjoint from $V(G)$,
for $1 \leq i \leq m$.
Attribute set $s_i$ of leaf vertex $v_i$ is $\{w: w\in V(G)$ and $wu_{i-m}\in E(G)\}$, 
for $m < i \leq 2m$.
}

\SubSection{A 2-Approximate Algorithm}
\vspace{-2pt}
\label{subsec:approxalgo}
An efficient dynamic programming based algorithm to find the optimal 
solution (under the benefit model presented in the previous section) 
for \refproblem{problem1} exists when the tree is a {\em path}. Note
that left-deep and right-deep join plans result in paths. 
\fullpaper{
We first present this algorithm and make use of it in finding a 
2-approximation for binary trees.
}
\shortpaper{
We present the solution for paths in brief and make use of the same 
in the 2-approximation for binary trees.
}

Consider a path $v_1, v_2, \ldots v_n$, where each vertex $v_i$ is associated 
with an attribute
set $s_i$. The optimal solution for
any segment $(i, j)$ of the path, {\small OPT$(i, j)$ = max$\{$ OPT$(i, k) + $OPT$(k+1, j) + c(i, j)$ $\}$} over all $i\leq k<j$, where $c(i, j)$ is the number of common 
attributes for the segment $(i, j)$. 
\shortpaper{
The detailed algorithm can be found in~\cite{FULLPAPER}.
}

\fullpaper{
The dynamic programming algorithm starts by finding the optimal solution 
for all segments of size 1, then of size 2 and so on. At any segment $(i, j)$, 
every possible way of splitting the segment into smaller segments is
considered and the solution for the smaller segments obtained from the
memo structure. The optimal solution for segment $(i, j)$ is then 
memoized. Procedure {\em PathOrder} in \reffig{fig:pathalgo} gives
the details.
}

\paper{
\fullpaper{
\begin{figure}[ht]
\begin{minipage}{0.5\textwidth}
\algsize{
\ordinalg{
{\bf Input:} \\
\>  s[n]  : array of attribute sets \\
{\bf Output:} \\
\>  p[n]  : array of permutations \\
{\bf Data Structures:} \\
\>  benefit[n][n], split[n][n] : arrays of integers \\
\>  commons[n][n] : array of sets \\
%\>  split[n][n]   : array of integers \\
\>  apermute(s)   : Function to give an arbitrary permutation of s \\

{\bf Procedure} PathOrder \\
BEGIN \\
\>    for i=1 to n \\
%\>    for i=1; i$\leq$n; i++) $\{$ \\
\>\>        benefit[i][i] = 0; commons[i][i] = s[i]; split[i][i] = -1; \\
%\>\>        commons[i][i] = s[i]; \\
%\>\>        split[i][i]   = -1; \\

\>   for j=1 to n-1 \\
%\>   for (j=1; j $\leq$ n-1; j++) $\{$ \\
\>\>        for i = 1 to n-j \\
%\>\>        for (i = 1; i $\leq$ n-j; i++) $\{$ \\
\>\>\>            Let k be the index such that i $\leq$ k $<$ (i+j) and \\
\>\>\>            benefit[i][k]+benefit[k+1][i+j] is maximum \\
\>\>\>            commons[i][i+j] = commons[i][k] $\cap$ commons[k+1][i+j]; \\
\>\>\>            benefit[i][i+j] = benefit[i][k] + benefit[k+1][i+j] + \\ 
\>\>\>\>\>\>                        $|$commons[i][i+j]$|$; \\
\>\>\>            split[i][i+j] = k; \\
%\>\>        $\}$ \\
%\>    $\}$ \\

\>    Call MakePermutation(1, n); \\
END PROC  \\
\vspace{0.5in}

{\bf Procedure} MakePermutation(i, j) \\
BEGIN \\
\>   if (i = j) \\
\>\>        p[i] = Append apermute(commons[i][i]) to p[i]; \\
\>\>        return; \\
   
\>    for k=i to j \\
%\>    for (k=i; k $\leq$j; k++) $\{$ \\
\>\>        p[k] = Append apermute(commons[i][j]) to p[k]; \\
\>\>        For all (i', j') $\neq$ (i, j)  \\
\>\>\>          commons[i'][j'] = commons[i'][j'] $-$ commons[i][j]; \\
\>\>        m = split[i][j]; \\
\>\>        MakePermutation(i, m); \\
\>\>        MakePermutation(m+1, j); \\
%\>   $\}$ \\
%$\}$\\
END PROC  \\
}
}
\end{minipage}
\caption{Optimal Orders for Path}
\label{fig:pathalgo}
\end{figure}

}
}
\techreport{
\begin{figure}[ht]
\begin{minipage}{0.5\textwidth}
\algsize{
\ordinalg{
{\bf Input:} \\
\>  s[n]  : array of attribute sets \\
{\bf Output:} \\
\>  p[n]  : array of permutations or orders \\
{\bf Data Structures:} \\
\>  benefit[n][n], split[n][n] : arrays of integers \\
\>  commons[n][n] : array of attribute sets \\
%\>  split[n][n]   : array of integers \\
\>  apermute(s)   : Function that returns an arbitrary permutation of attribute set s \\

{\bf Procedure} PathOrder \\
BEGIN \\
\>    for i=1 to n \\
%\>    for i=1; i$\leq$n; i++) $\{$ \\
\>\>        benefit[i][i] = 0; commons[i][i] = s[i]; split[i][i] = -1; \\
%\>\>        commons[i][i] = s[i]; \\
%\>\>        split[i][i]   = -1; \\

\>   for j=1 to n-1 \\
%\>   for (j=1; j $\leq$ n-1; j++) $\{$ \\
\>\>        for i = 1 to n-j \\
%\>\>        for (i = 1; i $\leq$ n-j; i++) $\{$ \\
\>\>\>            Let k be the index such that i $\leq$ k $<$ (i+j) and 
                  benefit[i][k]+benefit[k+1][i+j] is maximum \\
\>\>\>            commons[i][i+j] = commons[i][k] $\cap$ commons[k+1][i+j]; \\
\>\>\>            benefit[i][i+j] = benefit[i][k] + benefit[k+1][i+j] + 
                                    $|$commons[i][i+j]$|$; \\
\>\>\>            split[i][i+j] = k; \\
%\>\>        $\}$ \\
%\>    $\}$ \\

\>    Call MakePermutation(1, n); \\
END PROC  \\
\vspace{0.5in}

{\bf Procedure} MakePermutation(i, j) \\
BEGIN \\
\>   if (i = j) \\
\>\>        p[i] = Append apermute(commons[i][i]) to p[i]; \\
\>\>        return; \\
   
\>    for k=i to j \\
%\>    for (k=i; k $\leq$j; k++) $\{$ \\
\>\>        p[k] = Append apermute(commons[i][j]) to p[k]; \\
\>\>        For all (i', j') $\neq$ (i, j)  \\
\>\>\>          commons[i'][j'] = commons[i'][j'] $-$ commons[i][j]; \\
\>\>        m = split[i][j]; \\
\>\>        MakePermutation(i, m); \\
\>\>        MakePermutation(m+1, j); \\
%\>   $\}$ \\
%$\}$\\
END PROC  \\
}
}
\end{minipage}
\caption{Optimal Orders for Path}
\label{fig:pathalgo}
\end{figure}

}

For binary trees we propose an approximation with benefit at least half 
that of an optimal solution. We split the tree into two sets of paths,
$P_o$ and $P_e$. $P_o$ has the paths formed by edges at odd levels 
and $P_e$ has those formed by edges at even levels, \reffig{fig:treeapprox} 
shows an example. 
We 
\fullpaper{use the 
procedure {\em PathOrder} to find optimal solutions for each of the two
sets of paths.
}
\shortpaper{
obtain an optimal solution for each of the two sets of paths.
}
Let the the optimal solutions for the two sets of paths be $S_o$ and $S_e$ 
and the corresponding benefits be {\em ben($S_{o}$)} and {\em ben($S_{e}$)}.
Let the set of edges included in $P_o$ and $P_e$ be denoted by $E_{o}$ and
$E_e$ respectively.  Consider an optimal 
solution {\em $S_T$} for the whole tree. In the optimal solution, let the sum 
of benefits of all edges in $E_{o}$ be {\em odd-ben($S_T$)} and that of edges in
$E_{e}$ be {\em even-ben($S_T$)}.
Note that  {\em ben($S_o$)} $\geq$ {\em odd-ben($S_T$)} and {\em ben($S_e$)} 
$\ge$ {\em even-ben($S_T$)}. Since the total benefit of the optimal solution 
{\em ben($S_T$)} $=$ {\em odd-ben($S_T$)} $+$ {\em even-ben($S_T$)}, we have 
{\em ben($S_o$)} $+$ {\em ben($S_e$)} $\geq$ {\em ben($S_T$)}.
Hence at least one of {\em ben($S_o$)} or {\em ben($S_e$)} is $\geq 1/2$ {\em ben($S_T$)}.
There may be vertices not included in the chosen solution, e.g., the
even level split in \reffig{fig:treeapprox} does not include the root and leaf 
nodes. For these left over vertices arbitrary permutations
can be chosen. 
%\An improvement over this approximation and a lower bound
%on approximation can be found in ... 

\paper{
\begin{figure} [t]
%\centerline{\psfig{file=fig/treeapprox.eps}}
\includegraphics[angle=0,width=3.2in]{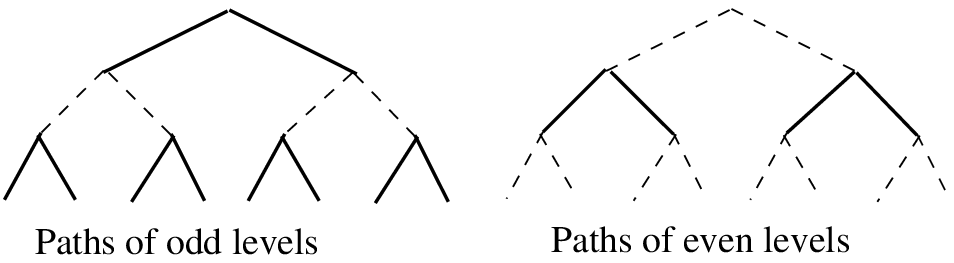}
\caption{A 2-approximation for binary trees}
\vspace{-3ex}
\label{fig:treeapprox}
\end{figure}
}

\techreport{
\begin{figure} [t]
\centerline{\psfig{file=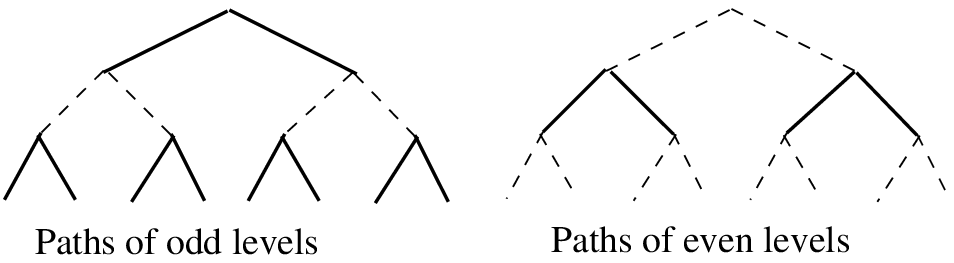}}
\caption{A 2-approximation for binary trees}
\label{fig:treeapprox}
\end{figure}
}

\Section {Optimization Exploiting Favorable Orders}
\label{sec:ford}
\eat{
As we showed in the previous section, choosing optimal sort orders 
is intractable even when the join order (logical rewriting) is fixed. 
Therefore, we propose a heuristic approach
that works in two phases. Below we briefly outline our approach
and give details in the rest of this section.
}
The benefit model we presented in the previous section and the approximation
algorithm do not take into account indices and size of relations or intermediate
results. Moreover, we assumed that the join order tree is fixed.
In this section we present a two phase approach for the more general
problem.
In phase-1, which occurs during plan generation, we exploit the 
information of available indices and properties of physical operators 
to efficiently compute a small set of promising sort orders to try. We 
formalize this idea through the notion of {\em favorable orders}.
% and
%describe how to compute a good approximation of {\em favorable 
%orders} efficiently. 
Phase-2, is a plan refinement step and occurs
after the optimizer makes its choice of the best plan. In phase-2,
the sort orders chosen by the optimizer are refined further to 
reap extra benefit from the attributes common to multiple joins.
Phase-2 uses the 2-approximate algorithm of \refsec{subsec:approxalgo}
\eat{
\noindent {\em Notations Used}:
{ \footnotesize
\begin{flushleft}
\begin{tabular}{|l|l|} \hline
{\em cbp}$(e, o)$   & Cost of the best plan for expression $e$ \\
                    & with $o$ being the required output order \\
\hline
$o_R$               & The clustering order of relation $R$ \\
\hline
{\em idx}$(R)$      & Set of all indices over $R$ \\
\hline
$o(I)$              & Order (key) of the index $I$ \\
\hline
$\langle s \rangle$ & An arbitrary order (permutation) on attribute set $s$ \\
\hline
$P(s)$              & Set of all permutations of set $s$ \\
\hline
$o \wedge s$        & Longest prefix of $o$ such that each attribute \\
                    & in the prefix belongs to the attribute set $s$ \\
\hline
\end{tabular}
\end{flushleft}
}
}

\eat{
\SubSection{Outline of the Approach}
In phase-1, which occurs during plan generation, we exploit the 
information of available indices and properties of physical operators 
to efficiently compute a small set of promising sort orders to try. We 
formalize this idea through the notion of {\em favorable orders} and
describe how to compute a good approximation of {\em favorable 
orders} efficiently. Phase-2, is a plan refinement step and occurs
after the optimizer makes its choice of the best plan. In phase-2,
the sort orders chosen by the optimizer are refined further to 
reap extra benefit from the attributes common to multiple joins.
Phase-2 uses the approximation algorithm of \refsec{subsec:approxalgo}
}

\eat{
Choice of good sort orders cannot be entirely pushed to a 
post-optimization phase as in~\cite{MITCH:VLDB03}. Not selecting
promising sort orders during optimization can lead to the pruning
of a sorting based plan, with the optimizer choosing an
alternative algorithm. The post-optimization phase will be too
late to recover from such sub-optimal choices made during
optimization. However, considering more number of interesting 
orders during 
optimization increases the optimization time. We adopt
a combined approach that avoids pruning of good sorting based
plans when the required order can be efficiently guaranteed and
also add a post-optimization refinement phase that allows us to
keep the overheads of optimization very low. We introduce the notion
of {\em favorable orders} that makes it possible to consider few
but important sort orders during the optimization phase.
}

\SubSection {Favorable Orders}
\label{subsec:favord}
%\vspace{1ex}

\eat{
\hrule
\vspace{0.5ex}
\noindent {\em Notations Used}:
\vspace{0.5ex}
\hrule
%{\small
\begin{itemize}
\addtolength{\itemsep}{-0.5pt}
\item {\em cbp}$(e, o)$ : Cost of the best plan for expression $e$
                     with $o$ being the required output order
\item $o_R$ : The clustering order of relation $R$ 
\item {\em idx}$(R)$ :  Set of all indices over $R$
\item $o(I)$ : Order (key) of the index $I$ 
\item $\langle s \rangle$ : An arbitrary order (permutation) on attribute set $s$
\item $P(s)$  : Set of all permutations of set $s$ 
\item $o \wedge s$ : Longest prefix of $o$ such that each attribute
in the prefix belongs to the attribute set $s$ 
\end{itemize}
%}
\hrule
\vspace{1ex}
}

Given an expression $e$, we expect some sort orders (on the result of
$e$) to be producible at much lesser cost than other sort orders. 
Available indices, indexed materialized views, specific rewriting 
of the expression 
and choice of physical operators determine what sort orders are easy 
to produce. To account for such orders, we introduce the notion of 
{\em favorable} orders. We use the following notations:

\vspace{-5pt}
\begin{itemize}
\addtolength{\itemsep}{-5pt}
\small{
\item {\em cbp}$(e, o)$ : Cost of the best plan for expression $e$
                     with $o$ being the required output order
\item $o_R$ : The clustering order of relation $R$ 
\item {\em idx}$(R)$ :  Set of all indices over $R$
\item $o(I)$ : Order (key) of the index $I$ 
\item $\langle s \rangle$ : An arbitrary order (permutation) on attribute set $s$
\item $P(s)$  : Set of all permutations of set $s$ 
\item $o \wedge s$ : Longest prefix of $o$ such that each attribute
in the prefix belongs to the attribute set $s$ 
\item {\em schema}$(e)$ : The set of attributes in the output of $e$ \\
}
\end{itemize}
\vspace{-8pt}

We first define the {\bf {\em benefit}} of a sort order $o$ {\em w.r.t.} 
an expression $e$ as follows:

\vspace{0.1in}
{\em benefit}$(o, e) = cbp(e, \epsilon)+coe(e, \epsilon, o)-cbp(e, o)$ \\

Intuitively, a positive benefit implies the order can be obtained with lesser cost
than a full sort of unordered result. For instance, the clustering order
of a relation $r$ will have a positive {\em benefit} for the expression 
$\sigma_{p}(r)$. 
Similarly, query covering secondary indices 
and indexed materialized views can yield orders with positive benefit.
We call the set of all orders, on $schema(e)$, having a positive 
benefit {\em w.r.t.} $e$ as the {\em favorable order set} of $e$ and denoted it
as {\em ford(e)}. 

\vspace{0.1in}
{\em ford(e)}$=\{$ {\em o: benefit(o, e)}$>0\ \}$ \\

\vspc{-6pt}
\subsubsection{Minimal Favorable Orders}

The number of favorable orders for an expression can be very large.
For instance, every order having the clustering order as its prefix is a favorable
order.
A {\em minimal favorable order set} of $e$, denoted by {\em ford-min(e)}, 
is the minimum size subset of {\em ford(e)} such that, for each order 
$o \in$ {\em ford(e)}, at least one of the following is true:

\begin{enumerate}
\addtolength{\itemsep}{-2mm}
    \item $o$ belongs to {\em ford-min(e)}
    \item $\exists$ $o'$ $\in$ {\em ford-min(e)} such that $o' \leq o$ and 
       {\em cbp}$(e, o')$ $+$ {\em coe}$(e, o', o)$ $=$ {\em cbp}$(e, o)$
        Intuitively, if the cost of obtaining order $o$ equals the cost
        of obtaining a partial sort order $o'$ followed by an explicit
        sort to get $o$, we include only $o'$ in the {\em ford-min}
    \item $\exists$ $o''$ $\in$ {\em ford-min(e)} such that $o \leq o''$ and
       {\em cbp}$(e, o'') =$ {\em cbp}$(e, o)$
        Intuitively, if an order $o''$ subsumes order $o$ and has the same
        cost, we include only $o''$ in {\em ford-min}
%    \item $\exists$ {\em \"{o}} $\in$ {\em ford-min(e)} such that {\em \"o} 
%        is a permutation of $o$ and {\em benefit(\"{o})} $\geq$ {\em benefit(o)}
\end{enumerate}

Conditions 2 and 3 above, ensure that when a relation has an
index on $o$, orders that are prefixes of $o$ and orders that have $o$ 
as their prefix are not included unnecessarily.

\noindent We define favorable orders of an expression {\em w.r.t.} a set 
of attributes $s$ as:
%\begin{center}
{\em ford(e, s)}$=\{o \wedge s$: $o\in$ {\em ford(e)}$\}$
%\end{center}
Intuitively, {\em ford(e, s)} is the set of orders on $s$ or a subset of $s$
that can be obtained efficiently.
Similarly, the {\em ford-min} of an expression {\em w.r.t.}
a set of attributes $s$ is defined as: 
%\begin{center}
{\em ford-min(e, s)}$=\{o \wedge s$ : $o\in$ {\em ford-min(e)}$\}$ \\
%\end{center}

\eat{
Before addressing the issue of computing the favorable orders for an 
expression, we describe how we compute the interesting orders for a 
join using the favorable orders. The technique is applicable to other
operators like Merge-Union as well. 
}

\vspc{-8pt}
\subsubsection{Heuristics for Favorable Orders}
\label{subsec:compfavord}
Note that the definition of favorable orders uses the cost of the best plan for
the expression. However, we need to compute the favorable orders of an 
expression {\bf before} the expression is optimized and without requiring to 
expand the logical or the physical plan space. Further, the size of the exact 
{\em ford-min} of an expression can be prohibitively large in the worst 
case.
%Fortunately, it is possible to 
%heuristically compute a good approximation of the favorable orders without
%optimizing the expression. 
\eat{
Since indices and physical operators (algorithms)
are the main factors that influence the favorable orders, it is possible 
to give good heuristics to compute the favorable orders approximately.
Note that we a priori know  the way each physical operator in the system 
propagates its input sort orders.
}
In this section, we describe a method of computing approximate 
{\em ford-min}, denoted as  {\em afm}, for {\em SPJG} expressions.
We compute the {\em afm} of an expression bottom-up. For any 
expression {\em e}, {\em afm(e)} is computable after the 
{\em afm}  is computed for all of {\em e}'s inputs.

\begin{enumerate}
\addtolength{\itemsep}{-1mm}
\item $e=R$, where $R$ is a base relation or materialized view.
We include the clustering order of $R$ and all secondary index orders such
that the index covers the query. \\
{\em afm}$(R)=\{o: o=o_R$ or $o=o(I), I \in idx(R)$ and $I$ covers the query$\}$
\item $e=\sigma_{p}(e_1)$, where $e_1$ is an arbitrary expression. \\
{\em afm}$(e)=\{o: o \in $ {\em afm($e_1$)} $\}$
\item $e=\Pi_{L}(e_1)$, where $e_1$ is any expression. We include longest prefixes 
of input favorable orders such that the prefix has only the projected attributes. \\
{\em afm}$(e)=\{o: \exists o'\in$ {\em afm($e_1$)} and $o=o'\wedge L\}$
\item $e=e_1 \Join e_2$ with join attribute set $S=\{a_1, a_2, \ldots a_n\}$. 
Noting that nested loops joins propagate the sort order of one of the inputs (outer) and 
merge join propagates the sort order chosen for join attributes, we compute
the {\em afm} as follows. First, we include all sort orders in the input {\em afms}. 
Next, we consider the longest prefix of each
input favorable order having attributes only from the join attribute set and extend
it to include an arbitrary permutation of the remaining join attributes.  \\
{\em afm}$(e_1 \Join e_2)=\mathcal{T}\bigcup\ \{o:o' \in \mathcal{T}\bigcup\{\epsilon\}\ and\ o=o' \wedge S\ + \ \langle S-${\em attrs}$(o'\wedge S)\rangle \}$, where $\mathcal{T}=${\em afm}$(e_1)\ \bigcup$ {\em afm}$(e_2)$ 

Note that, for the join attributes not involved in an input favorable order prefix
({\em i.e.,} $S-${\em attrs}$(o' \wedge S)$), we take an arbitrary permutation. An
exact {\em ford-min} would require us to include all permutations of such attributes.
%In the extreme
%case when there are no input favorable orders, we just take an arbitrary permutation
%of all the join attributes. 
In the post-optimization phase, we refine the choice made here using the benefit
model and algorithm of \refsec{subsec:approxalgo}.

%Note that for a complete {\em ford-min} we must 
%include {\em all} permutations of $S-attrs(o')$. Instead, we include 
%only one permutation and rework the permutation during the post-processing
%phase.

\item $e=_{L}\mathcal{G}_{F}(e_1)$ \\
%Let $L'$ be the set of group-by attributes 
%included in the output ({\em i.e.,} $L'=L\cap F$) \\
{\em afm}$(e)=$ 
          $\{o:o' \in $ {\em afm($e_1$)}$\bigcup \{\epsilon\}$ {\em and } $o=o' \wedge L\ + \ \langle L-${\em attrs}$(o'\wedge L)\rangle \}$\\
Intuitively, for each input favorable order we identify the longest prefix with
attributes from the projected group-by columns and extend the prefix with
an arbitrary permutation of the remaining attributes.
\end{enumerate}

Computing {\em afms} requires a single pass of the query tree. At each node of
the query tree the only significant operation performed is computation of the
longest common prefix {\em lcp} of an order {\em w.r.t.} a set, {\em i.e.,} $o \wedge s$. 
The number of {\em lcp} operations performed at a node is at most the number 
of input favorable orders for that node. Let $m$ be the average number of 
favorable orders for base relations. If the query involves a join of $n$ relations 
the worst-case number of {\em lcp} operations performed at any node is $O(2^n m)$. 
Typically $m$ is very small ($\leq 2$) and hence even the worst-case number
of {\em lcp} operations is within acceptable limits. However, in most practical cases, 
the number of {\em lcp} operations performed is far below the worst-case.

\eat{
\SubSection{Optimizer Changes}
\label{subsec:optchange}
The changes to be made for a cost-based optimizer to incorporate our techniques
are fairly simple. While considering a sorting based plan for a logical operator, 
we compute the set $\mathcal{I}(e,o)$ as explained
above and optimize inputs for each order in $\mathcal{I}(e,o)$. Our 
experiments show that the size of the set $\mathcal{I}(e,o)$ is generally
small and therefore the overheads of optimization are very less.
}

\eat{
The {\em Volcano} optimizer assumes that an algorithm (physical operator) either 
guarantees a required sort order fully or it does not. However, it is possible that 
an algorithm satisfies the required output order partly. For examples, the
required output order on the output of a merge join could be $(a,b,c,d)$ where
as the join predicate could be on attributes $\{a, b, e, f\}$. The merge-join
could obtain its inputs sorted on $(a, b, e, f)$ and have an enforcer above,
which enforces $(a, b, c, d)$ knowing that its input guarantees the prefix
$(a, b)$. For an optimization goal of (expression $e$, output order $o$),
the optimizer should consider all physical operators such that an order
$o' \leq o$ is guaranteed by the operator. Accordingly, the cost of a 
partial enforcer $coe(e, o', o)$ is to be added.
}

\SubSection {Overall Optimizer Extensions}
\label{subsec:optext}
We make use of the approximate favorable orders during plan generation (phase-1) 
to choose a small set of promising {\em interesting orders} for sort-based 
operators. We describe our approach taking merge join as an example but the approach
is applicable to other sort based operators. In phase-2, which is a post-optimization
phase, we further refine the chosen sort orders.

%\subsection {Interesting Orders for Merge-Join}

\vspc{-6pt}
\subsubsection{Plan Generation (Phase-1)}
Consider an optimization goal of expression $e=e_l \Join e_r$ and required
output sort order $o$. When we consider merge-join as a candidate algorithm, we
need to generate sub-goals for $e_l$ and $e_r$ with the required output
sort order being some permutation of the join attributes. 
%First, we 
%define a set $\mathcal{I}(e, o)$, that is {\em guaranteed} to contain an optimal sort 
%order for the merge-join plan. 

Let $S$ be the set of attributes involved in the join predicate. We consider 
only conjunctive and equality predicates.  We compute the set $\mathcal{I}(e, o)$ 
of interesting orders as follows:

%We are now ready to define the set $\mathcal{I}(e)$ of interesting orders for 
%a join expression $e=e_l\Join e_r$ with $o$ as the required order on the result.
%Let $S$ be the set of attributes involved in the 
%join predicate. We consider only conjunctive and equality predicates.
%The set of interesting orders $\mathcal{I}(e)$ is then computed as follows: \\

\begin{enumerate}
\addtolength{\itemsep}{-0.3pt}
    \item Collect the favorable orders of inputs plus the required output order \\
        $\mathcal{T}(e, o) = ${\em afm}$(e_l, S)\ \bigcup\ ${\em afm}$(e_r, S)\ \bigcup\ {o \wedge S}$, where {\em afm}$(e, S)=\{o' \wedge S: o'\in $ {\em afm}$(e)\}$ 

    \item Remove redundant orders \\
          If $o_1, o_2 \in \mathcal{T}(e,o)$ and $o_1 \leq o_2$, remove $o_1$ from $\mathcal{T}(e,o)$
    \item Compute the set $\mathcal{I}(e,o)$ by extending each order in 
          $\mathcal{T}(e,o)$ to the length of $|S|$; the order of extra 
         attributes can be arbitrarily chosen \\
 $\mathcal{I}(e,o)=\{o : o'\in \mathcal{T}(e,o)\ and\ o=o'\ + \ \langle S-attrs(o')\rangle \}$

\eat{
    \item Compute the set $\mathcal{I}(e,o)$ by extending each order in $\mathcal{T}(e,o)$ to the length of $|S|$ in {\em every} possible way \\
% $\mathcal{I}(e,o)=\{o : o'\in \mathcal{T}(e,o)\ and\ o=o'\ + \ \langle S-attrs(o')\rangle \}$
{\small
$\mathcal{I}(e,o)=\{o : \exists o'\in \mathcal{T}(e,o),\ \exists o''\in P(S-${\em attrs}$(o'))\ and\ o=o'\ + \ o''\}$
}
%\hl{ $\mathcal{I}(e,o)=\{o : \exists o'\in \mathcal{T}(e,o),\ $} \\
%\hl{$\exists o''\in P(S-attrs(o'))\ and\ o=o'\ + \ o''\}$}
}
\end{enumerate}

%Once the set $\mathcal{I}(e,o)$ is computed, 
We then generate optimization sub-goals 
for $e_l$ and $e_r$ with each order $o'\in\mathcal{I}(e,o)$ as the required 
output order and retain the cheapest combination.

\noindent {\bf A Note on Optimality:}
If the set $\mathcal{I}(e,o)$ is computed using the exact {\em ford-mins}
instead of {\em afms}, we claim that it must contain an optimal sort order 
(a sort order that produces the optimal merge join plan in terms of overall 
plan cost). 
\shortpaper{
The detailed proof of this claim can be found in our full paper~\cite{FULLPAPER}.
}
\fullpaper{
Appendix \ref{sec:proof} gives the detailed proof of this claim.
}

\eat{
\noindent
\begin{claim}
\label{claim1}
The set $\mathcal{I}(e,o)$ contains an {\em optimal order} $o_p$ for
the optimization goal ${\mathcal G}$, under the assumption (A)

\noindent Assumption (A): \\
If $o_1$, $o_2$ are two orders such that {\em attrs}$(o_1)=${\em attrs}$(o_2)$, 
cpu-cost$(e, o_1)=$cpu-cost$(e, o_2)$~\symbolfootnote[1]{Though this is not a very accurate cost model, we restrict ourselves to this model 
as the problem is hard even for this simple model. 
}
}

\eat{
\noindent
%For a formal proof of Claim~\ref{claim1} refer to \refsec{sec:proof}.
{\em For a formal proof of the claim we refer to the full length paper~\cite{FULLPAPER}}
\end{claim}

% in the extended version of the paper.
There are two main difficulties in computing and using the set $\mathcal{I}(e,o)$.
\begin{enumerate}
\item In the worst case, the size of set $\mathcal{I}(e,o)$ can be same as 
      the set of all permutations (see step 3 in its deduction).
\item Computing $\mathcal{I}(e,o)$  requires accurate {\em ford-mins} for 
      the input expressions, which is not possible unless the input plan
      space is explored fully. 
\end{enumerate}
We therefore provide (in \refsec{subsec:compfavord}) a way of
computing a reasonable approximation of {\em ford-mins} and hence the set
$\mathcal{I'}(e,o)$, an approximation of $\mathcal{I}(e,o)$. 
}

\noindent {\bf An Example:}
%We illustrate our technique with an example.
Consider \refexample{example1} of \refsec{sec:pso}. For brevity, we refer to
the two catalog tables as {\em ct1} and {\em ct2}, the rating table as {\em rt} 
and the columns with their starting letters.
The {\em afms} computed as described in \refsec{subsec:compfavord} are as 
follows:
{\em afm}$(ct1)=\{(y)\}$, {\em afm}$(ct2)=\{(m)\}$, {\em afm}$(rt)=\{(m)\}$,
{\em afm}$(ct1 \Join ct2)=\{(y,co,c,m),(m,co,c,y)\}$, 
{\em afm}$((ct1 \Join ct2) \Join rt)=\{(y,m), (m,y)\}$

For $(ct1 \Join ct2) \Join rt$ we consider two interesting sort orders $\{(y,m), (m,y)\}$
and for $ct1 \Join ct2$ we consider the four orders $\{(y,co,c,m),(m,co,c,y),(y,m,co,c),
(m,y,co,c)\}$. As a result the optimizer arrives at plan shown in \reffig{fig:optplan}.

\eat{
 but fails to notice that, for
$ct1 \Join ct2$, 
choosing the order $(m,y,co,c)$  instead of $(m,y,c,co)$ would have been better since
this order would have got passed through the join above and the cost of the top-most
sort enforcer (3) would have been lowered. We call this, the problem of choosing 
globally optimal interesting order. As we show in the next section selecting a globally 
sort order turns out to be NP-Hard. As part of our post-optimization order 
refinement phase, we adopt a 2-approximation to rectify the choices made
during optimization.
}

\vspc{-6pt}
\subsubsection{Plan Refinement (Phase-2)}
During the plan refinement phase, for each merge-join node in the plan tree,
we identify the set of {\em free attributes}, the attributes which were not
part of any of the input favorable orders. Note that for these attributes we had chosen
an arbitrary permutation while computing the {\em afm} (\refsec{subsec:compfavord}).
We then make use of the 2-approximate algorithm for trees 
(\refsec{subsec:approxalgo}) and rework the permutations
chosen for the {\em free attributes}. 

Formally, let $p_i$ be the permutation 
chosen for the join node $v_i$. Let $q_i$ be the order such that $q_i \in $ 
{\em afm}$(v_i.${\em left-input}$)$ $\cup$ {\em afm}$(v_i.${\em right-input}$)$
and 
$|p_i \wedge q_i|$ is the maximum. Intuitively, $q_i$ is the input favorable
order sharing the longest common prefix with $p_i$.
%the chosen permutation for $v_i$.
Let $f_i=${\em attrs}$(p_i - (p_i \wedge q_i))$; $f_i$ is the set of {\em free attributes}
for $v_i$. 

We now construct a binary tree where each node $n_i$ corresponding
to join-node $v_i$ is associated with the attribute set $f_i$. The orders 
for the nodes are chosen using the 2-approximate 
algorithm; the chosen order for free attributes is then appended
to the order chosen during plan generation ({\em i.e.,} $p_i \wedge q_i$)
to get a complete order.

%the algorithm described in \refsec{subsec:approxalgo}. 
The reworking of the orders will be
useful only if the adjacent nodes share the same prefix, i.e., $p_i \wedge q_i$
was the same for adjacent nodes. This condition however certainly holds when 
the inputs for joins have no favorable orders.
% and the join attributes have overlap.

\reffig{fig:postpassexample} illustrates the post-optimization phase. Assume
all relations involved ($R_1\ldots R_4$) are clustered on attribute $a$ and
no other favorable orders exist. {\em i.e.,} {\em afm}$(R_i)=\{(a)\}$, 
for $i=1$ to $4$. The orders chosen by the plan generation phase are shown besides the 
join nodes with {\em free attributes} being in {\em italics}. The reworked
orders after the post-optimization phase are shown underlined.

\paper{
\begin{figure} [t]
\centerline{\psfig{file=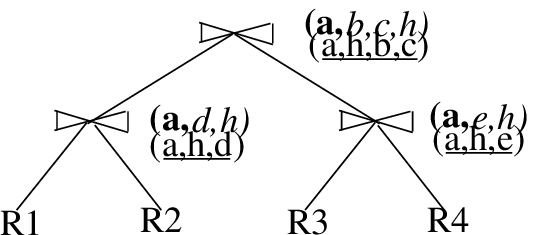}}
\caption{Post-Optimization Phase}
\label{fig:postpassexample}
\end{figure}
}

\techreport{
\begin{figure} [ht]
%\centerline{\psfig{file=fig/postpassexample.eps}}
\centerline{
\includegraphics[angle=0,width=3.5in]{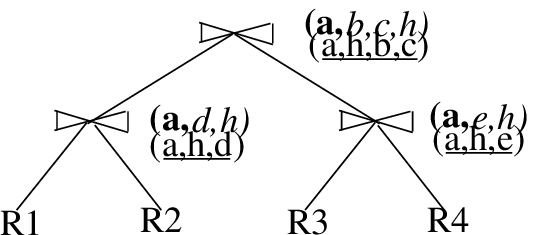}
}
\caption{Post-Optimization Phase}
\label{fig:postpassexample}
\end{figure}
}

\Section{Experimental Results} 
\label{sec:perf}
We performed experiments to evaluate the benefits our techniques. 
%In the second set of experiments 
%we evaluate our approach of choosing interesting orders so as to 
%exploit complete or partial sort orders available from input 
%subexpressions. 
For comparison, we use PostgreSQL (version 8.1.3) 
and two widely used commercial database 
systems (we call them SYS1 and SYS2). All tests were run on an Intel P4 (HT) 
PC with 512 MB of RAM. We used TPC-H 1GB dataset and additional tables as
specified in the individual test cases. For each table, a clustering index 
was built on the primary key. Additional secondary indices built are specified 
along with the test cases. All relevant statistics were built and the 
optimization level for one of the systems, which supports multiple levels of 
optimization, was set to the highest.

\eat{
The tables used and the number of tuples in each table are summarized
in \reffig{table:perftables}.

\begin{figure}[t]
\center
{ \small
\begin{tabular}{|c|c|c|c|} \hline
{\bf Name} & lineitem & part & supplier \\
\hline
{\bf Tuples} & 6,000,000 & 200,000 & 10,000 \\
\hline
\hline
{\bf Name} & partsupp & orders & customer\\
\hline
{\bf Tuples} &  800,000 & 1,500,000 &  150,000 \\
\hline
\end {tabular}
}
\caption{Tables used for the Experiments}
\label{table:perftables}
\end{figure}
}

%We used the TPC-H dataset of 1 GB for our tests. 

\SubSection{Modified Replacement Selection}
The first set of experiments evaluate the benefits of modified replacement 
selection (MRS) as compared to the standard replacement selection (SRS) 
when the input is known to be partially sorted. 

External sort in PostgreSQL employs the standard replacement selection (SRS)
algorithm suitably adapted for variable length records.  We modified 
this implementation to exploit partial sort orders available on the input.
%For the two commercial systems, SYS1 and SYS2, we did not have access
%to the source code and hence simulated the effect 
%of MRS by using nested rank queries. With the nested rank 
%queries we could independently sort multiple groups of tuples, each 
%group having the set of tuples with the same value for attributes in 
%the partial sort order already available on the input.

\noindent {\bf Experiment A1}: The first experiment consists of a simple 
ORDER BY of the TPC-H {\em lineitem} table on two columns {\em (l$\_$suppkey, 
l$\_$partkey)}. 

\paper{
\newtheorem{query}{Query}
\begin{query}
ORDER-BY on lineitem
\label{query:lisort}
\vspace{-5pt}
{\footnotesize
\begin{tabbing}
xxxxxxxxxx\=xxxxxxxxx\=xxxxxxxxx\=\kill
SELECT l$\_$suppkey, l$\_$partkey FROM lineitem \\
ORDER BY l$\_$suppkey, l$\_$partkey;
\vspace{-5pt}
\end{tabbing}
}
\end{query}
}

\techreport{
\newtheorem{query}{Query}
\begin{query}
ORDER-BY on lineitem
\label{query:lisort}
\vspace{-5pt}
{\footnotesize
\begin{tabbing}
xxxxxxxxxx\=xxxxxxxxx\=xxxxxxxxx\=\kill
SELECT l$\_$suppkey, l$\_$partkey FROM lineitem 
ORDER BY l$\_$suppkey, l$\_$partkey;
\end{tabbing}
}
\end{query}
}

A secondary index on {\em l$\_$suppkey} was available that 
covered the query (included the {\em l$\_$partkey} column)\footnote{On systems
not supporting indexes with included columns, we used a table with only the
desired two columns, clustered on {\em l$\_$suppkey}}.
On all three systems, the order by on {\em (l$\_$suppkey, l$\_$partkey)}
took almost the same time as an order 
by on {\em (l$\_$partkey, l$\_$suppkey)} showing that the sort operator 
of these systems did not exploit partial sort orders effectively. We 
then compared
the running times with our implementation that exploited partial sort
order {\em (l$\_$suppkey)} and the results are shown 
in \reffig{fig:perf-pso}. 

On SYS1 and SYS2 we simulated the
partial sorting using a correlated rank query (as we did not have access
to their source code). The subquery sorted the 
index entries matching a given {\em l$\_$suppkey} on {\em l$\_$partkey} 
and the subquery 
was invoked with all {\em suppkey} values so as to obtain the desired
sort order of {\em (l$\_$suppkey, l$\_$partkey)}. 
%Further, on all three systems (with their default sort), an order by on 
%{\em (l$\_$partkey, l$\_$suppkey)} took almost the same time as an order 
%by on {\em (l$\_$suppkey, l$\_$partkey)} showing that the systems by
%default did not exploit partial sort orders effectively.

\paper{
\fullpaper{
\begin{figure} [t]
\includegraphics[angle=270,width=3.3in]{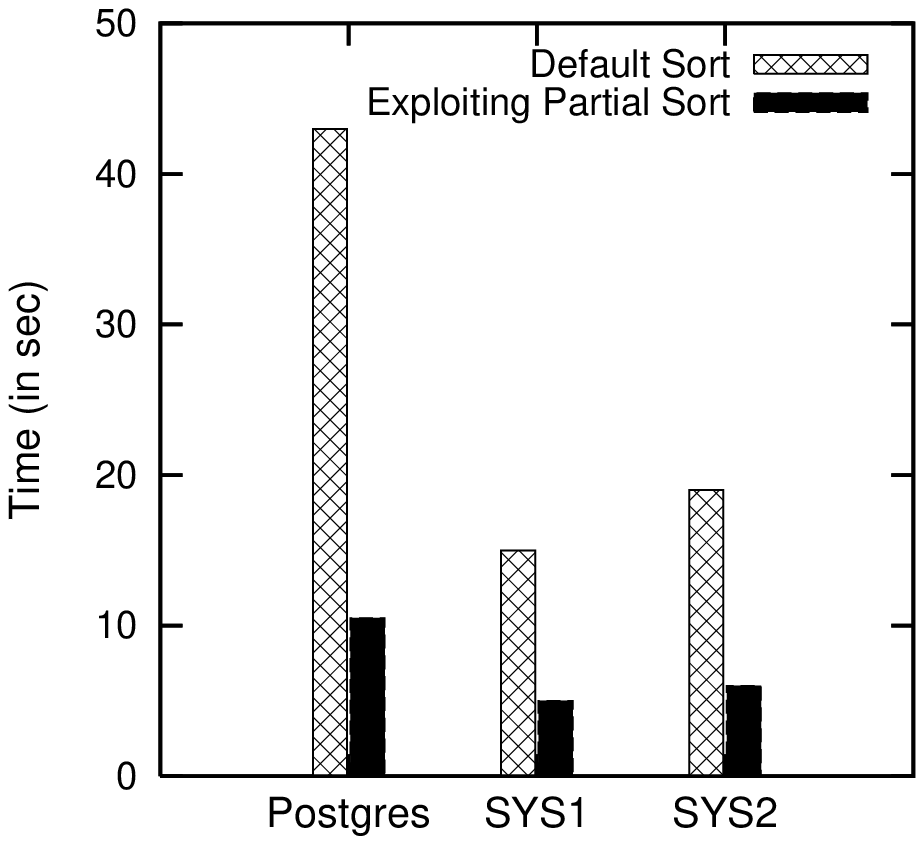}
%\caption{Gains of Exploiting Partial Sort Orders} 
\caption{Gains of Partial Sorting} 
\label{fig:perf-pso}
\end{figure}
}
}

\techreport{
\begin{figure}[ht]
%\centering
\hspace{-3ex}
\begin{tabular}{@{}c@{}c}
\begin{minipage}{0.5\textwidth} %DEFINES THE WIDTH OF EACH PICTURE IN % OF PAGE WIDTH
\includegraphics[angle=270,totalheight=2.2in,width=4.0in]{perf/plots/lisort.eps}
%\caption{Gains of Partial Sorting} 
\caption{Experiment 1} 
\label{fig:perf-pso}
\end{minipage}
&
\begin{minipage}{0.5\textwidth} %DEFINES THE WIDTH OF EACH PICTURE IN % OF PAGE WIDTH
\includegraphics[angle=270,totalheight=2.2in,width=4.0in]{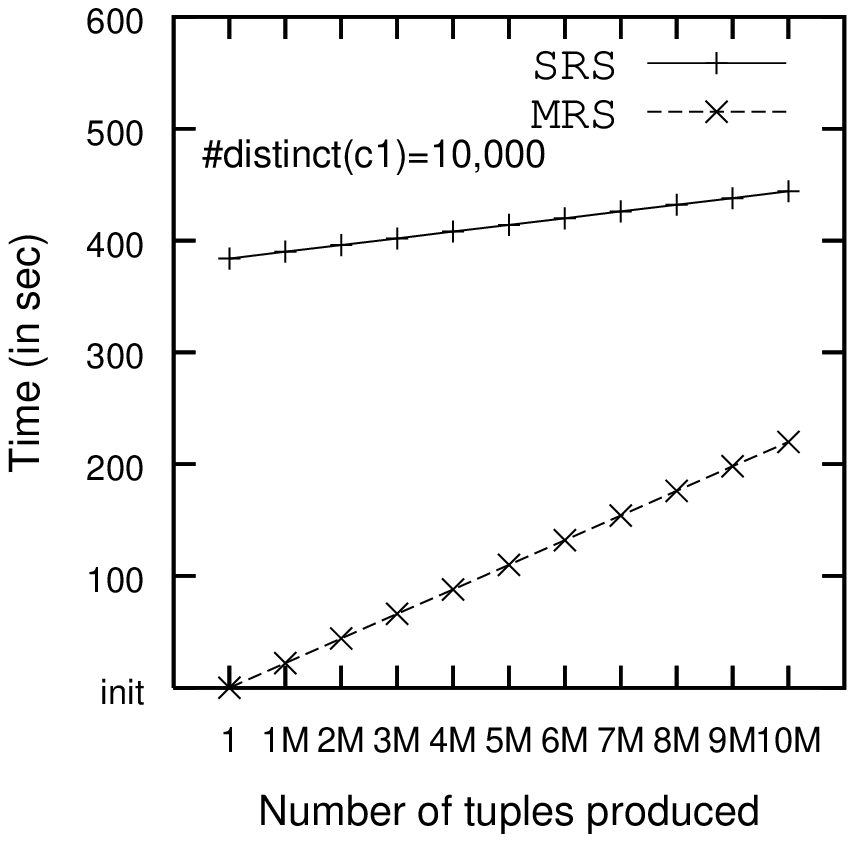}
\caption{Rate of Output}
\label{fig:perf-rate}
\end{minipage}
\end{tabular}
\end{figure}
}

\shortpaper{
\begin{figure}[t]
%\centering
\hspace{-3ex}
\begin{tabular}{@{}c@{}c}
\begin{minipage}{0.25\textwidth} %DEFINES THE WIDTH OF EACH PICTURE IN % OF PAGE WIDTH
\includegraphics[angle=270,totalheight=1.5in,width=2.2in]{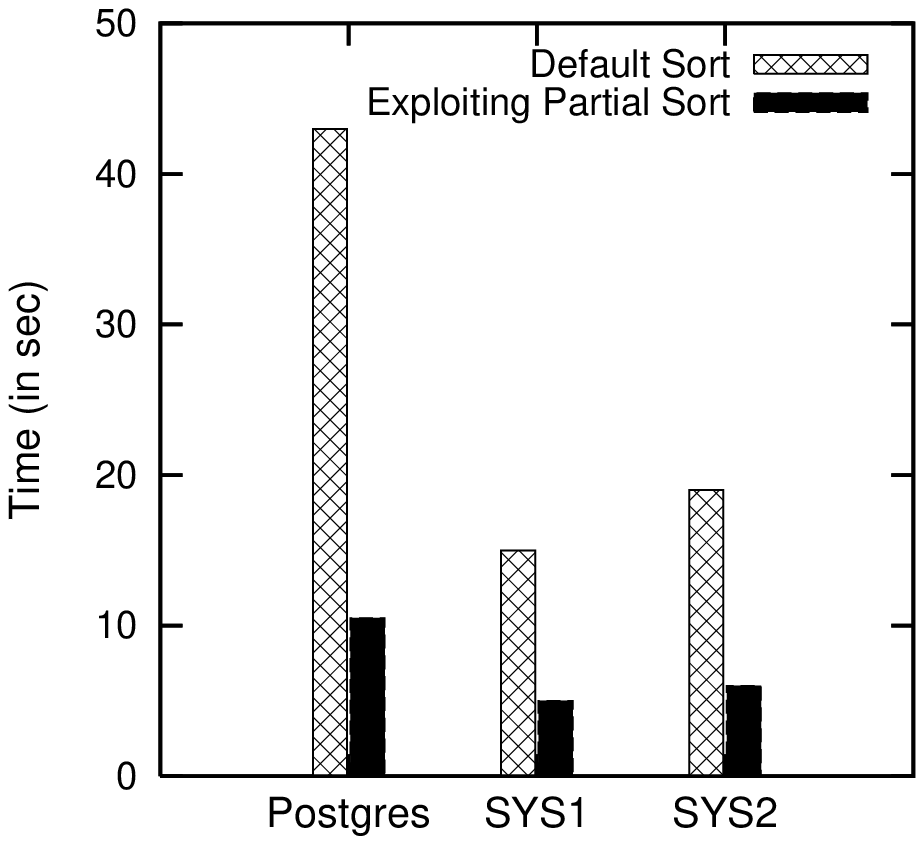}
%\caption{Gains of Partial Sorting} 
\caption{Experiment 1} 
\label{fig:perf-pso}
\end{minipage}
&
\begin{minipage}{0.25\textwidth} %DEFINES THE WIDTH OF EACH PICTURE IN % OF PAGE WIDTH
\includegraphics[angle=270,totalheight=1.6in,width=2.2in]{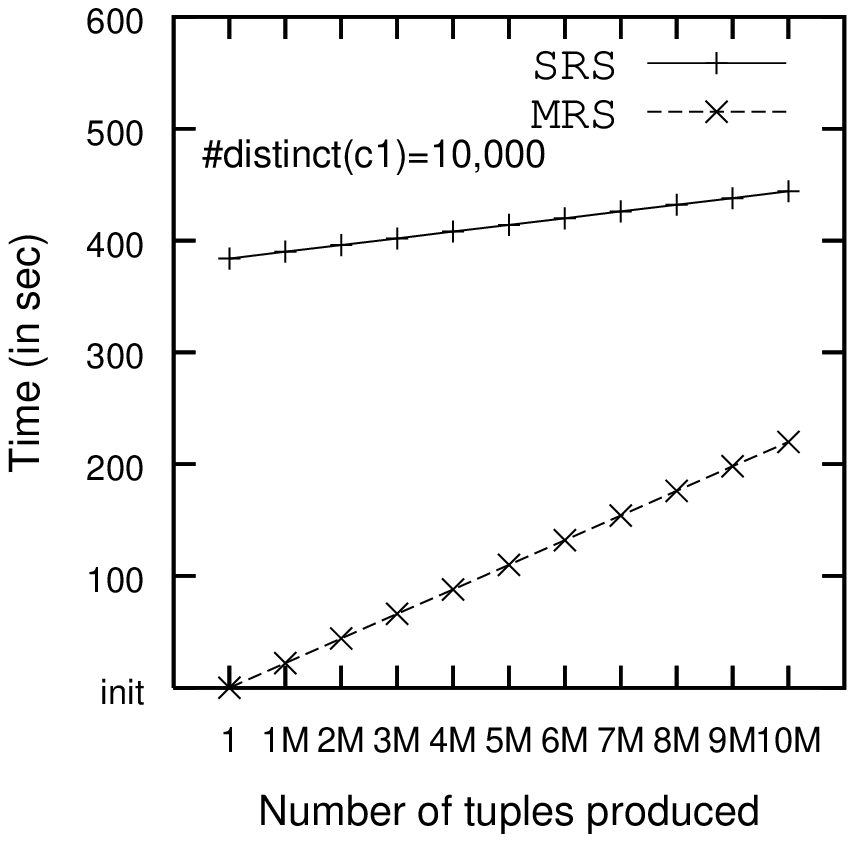}
\caption{Rate of Output}
\label{fig:perf-rate}
\end{minipage}
\end{tabular}
\end{figure}
}

By avoiding run generation I/O and making reduced comparisons, MRS performs 3-4 
times better than SRS. 

\noindent {\bf Experiment A2}: The second experiment shows how MRS 
is superior in terms of its ability to produce records 
early and uniformly. Table $R_3$ having 3 columns $(c1, c2, c3)$ was
populated with 10 million records and was clustered on $(c1)$.
The query asked an order by on $(c1, c2)$. \reffig{fig:perf-rate} shows
the plot of number of tuples produced vs. time with cardinality 
of $c1=10,000$.

\paper{
\fullpaper{
\begin{figure} [t]
\includegraphics[angle=270,width=3.3in]{perf/plots/rate_time_y.eps.big}
\caption{Rate of Output} 
\label{fig:perf-rate}
\end{figure}
}
}

MRS starts producing the tuples without any delay after the operator initialization
where as SRS produces its first output tuple only after seeing all input tuples.
By producing tuples early, MRS speeds up the pipeline significantly and also helps
Top-K queries.

\noindent {\bf Experiment A3}: The third experiment shows the effect of 
{\em partial sort segment size} on sorting. 8 tables $R_0 \ldots R_7$,
with identical schema of 3 columns $(c1, c2, c3)$ were each populated 
with 10 million
records and average record size of 200 bytes.  Each table was clustered
on $(c1)$. Table $R_{i}$ had $10^i$ tuples for each value of $c1$, resulting
in a partial sort segment size of  
$200\times 10^i$ bytes. Thus
$R_0$ had $c1$ as unique and sort segment size of 200 bytes 
and $R_7$ had the same value of $c1$ for all
10 million records leading to a sort segment size of 2GB. 
The query asked for an order by on $(c1, c2)$.
The running times with default and modified replacement
selection on PostgreSQL are shown in \reffig{fig:perf-pgvaryselect}.

\paper{
\begin{figure} [t]
\includegraphics[angle=270,width=3.3in]{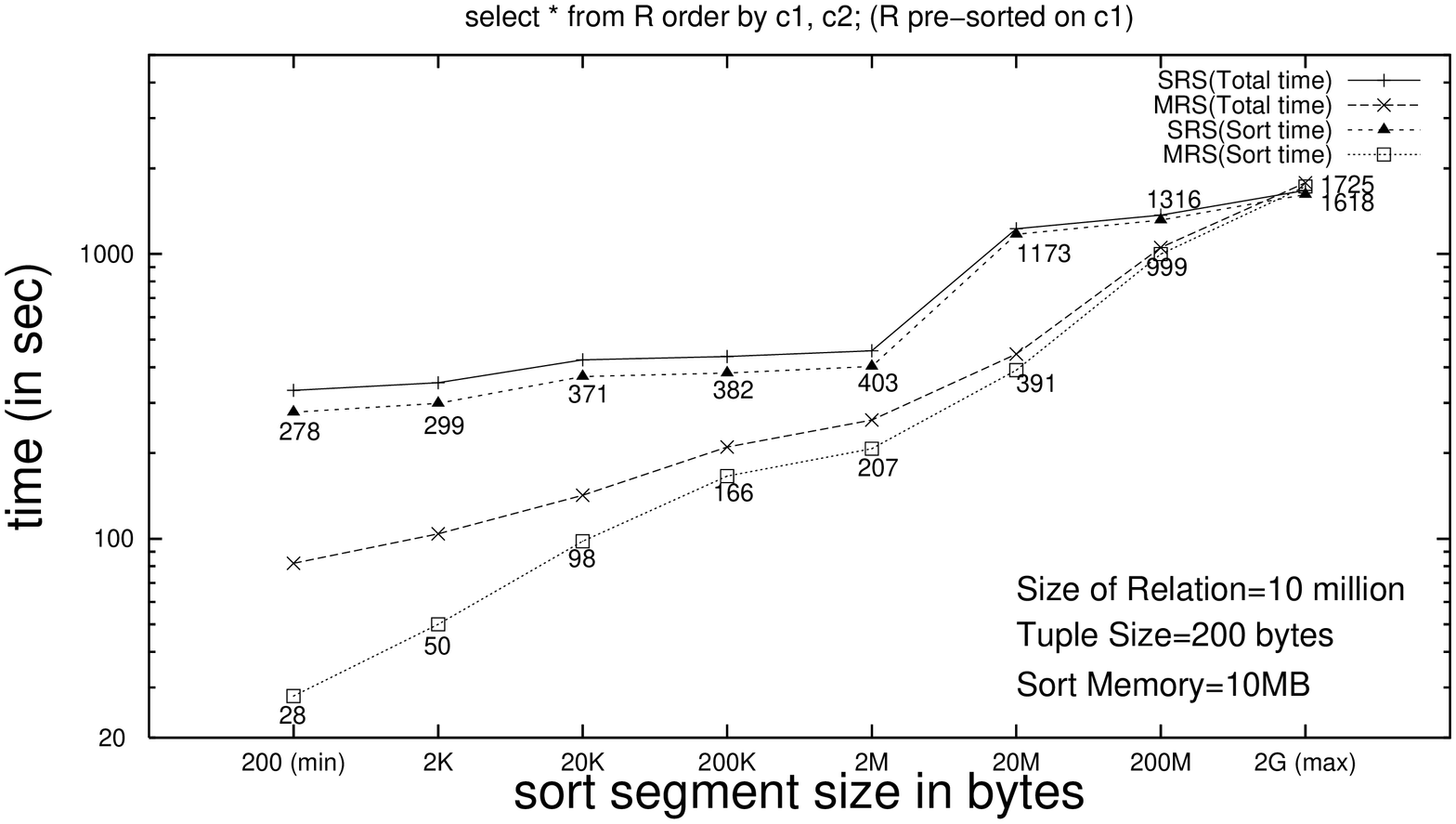}
\caption{Effect of Partial Sort Segment Size}
\label{fig:perf-pgvaryselect}
\end{figure}
}

\techreport{
\begin{figure} [t]
\includegraphics[angle=270,width=5.0in]{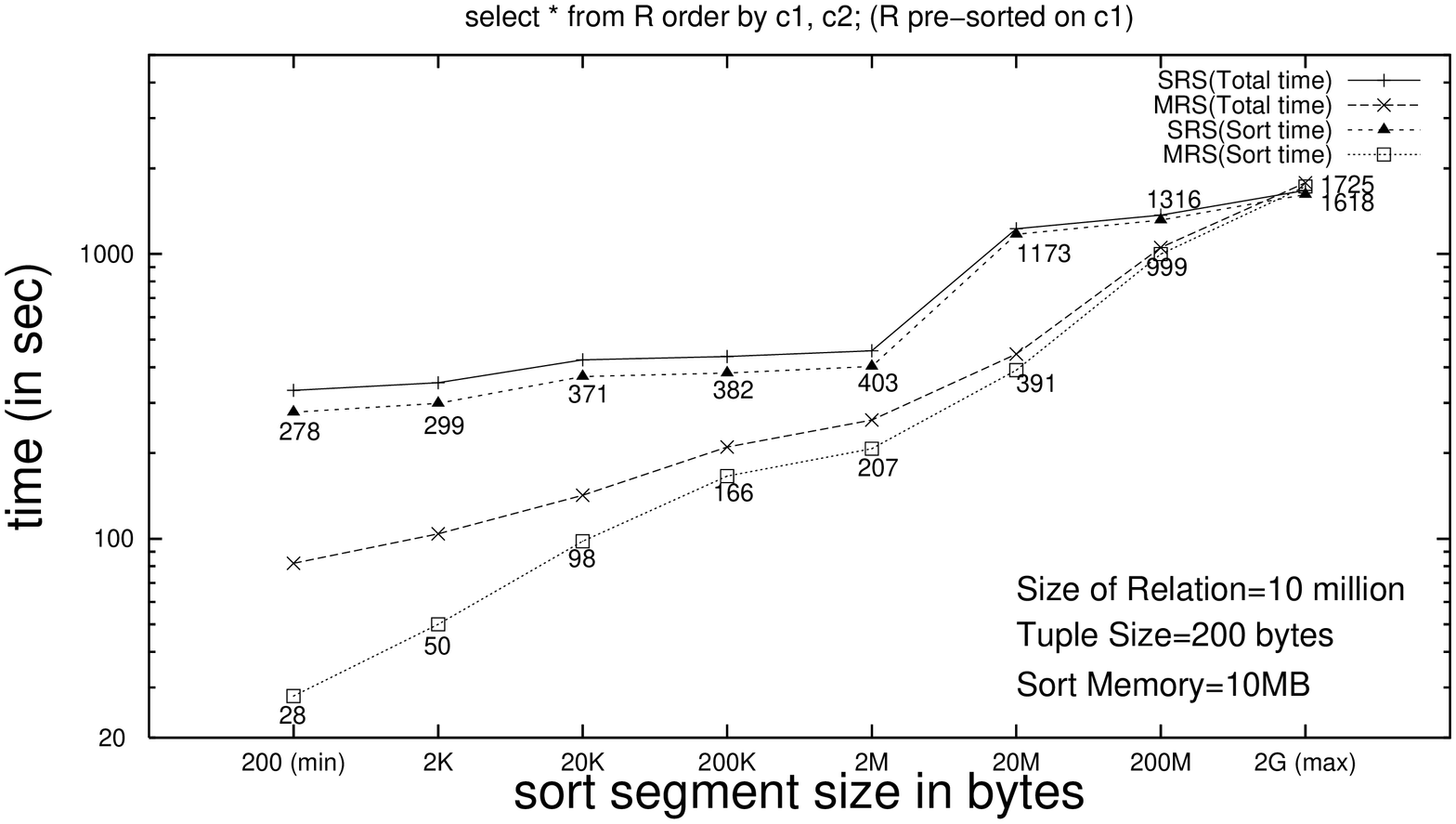}
\caption{Effect of Partial Sort Segment Size}
\label{fig:perf-pgvaryselect}
\end{figure}
}

When the partial sort segment size is small enough to fit in memory (up to
10MB or 50K records), SRS produces a 
single sorted run on disk and does not involve merging of runs. The modified
replacement selection (MRS) gets the benefit of avoiding I/O and reduced number
of comparisons. When the partial sort segment size becomes too large to 
fit in memory, we see a sudden rise in the time taken by SRS.  This is 
because replacement selection will have to deal
with merging several runs. MRS however deals with 
merging smaller number of runs initially as each partial sort segment is 
sorted separately. As the partial sort segment size increases, the running
time of MRS rises and becomes same as that of SRS
at the extreme point where all records have
the same value for $c1$.

\noindent {\bf Experiment A4}: To see the influence of MRS on a query having
other operators, we considered a query that asked for counting the number of
lineitems for each supplier, part pair. 
Two indices, {\em lineitem(l$\_$suppkey)} and  {\em partsupp(ps$\_$suppkey)},
each of which included other required columns supplied the required 
sort order partially.
%The {\em partsupp} table was clusterd on
%{\em ps$\_$suppkey} and a secondary index {\em lineitem(l$\_$suppkey)} that 
%covered the query supplied the partial sort order.

\begin{query}
\label{query:pslijoin}
Number of lineitems for each (supplier, part) pair
{\footnotesize
\begin{tabbing}
xxxxxxxxxx\=xxxxxxxxx\=xxxxxxxxx\=\kill
SELECT \> ps$\_$suppkey, ps$\_$partkey, ps$\_$availqty, count(l$\_$partkey) \\
FROM   \> partsupp, lineitem \\
WHERE  \> ps$\_$suppkey=l$\_$suppkey AND ps$\_$partkey=l$\_$partkey \\
xxxxxxxxxxxx\=xxxxxxxxx\=xxxxxxxxx\=\kill
GROUP BY \> ps$\_$suppkey, ps$\_$partkey, ps$\_$availqty \\
ORDER BY \> ps$\_$suppkey, ps$\_$partkey;
\end{tabbing}
}
\end{query}

\eat{
\noindent {\bf Query 2:} Number of lineitems for each supplier, part pair
{\em 
{\footnotesize
\begin{tabbing}
xxxxxxxx\=xxxxxxxxx\=xxxxxxxxx\=\kill
SELECT \> ps$\_$suppkey, ps$\_$partkey, count(l$\_$partkey) \\
FROM   \> partsupp, lineitem \\
WHERE  \> ps$\_$suppkey=l$\_$suppkey AND ps$\_$partkey=l$\_$partkey \\
xxxxxxxxxxxx\=xxxxxxxxx\=xxxxxxxxx\=\kill
GROUP BY \> ps$\_$suppkey, ps$\_$partkey \\
ORDER BY \> ps$\_$suppkey, ps$\_$partkey;
\end{tabbing}
}
}
}

The query took 63 seconds to execute with SRS and 25 seconds with MRS, both
on Postgres.
%\reffig{fig:perf-pslijoin} shows the running times of the query with SRS and MRS
%on Postgres. 
The query plan used in both cases was the same - a merge join of the
two relations on {\em (suppkey, partkey)} followed by an aggregate.

\eat{
\begin{figure} [t]
\includegraphics[angle=270,width=3.3in]{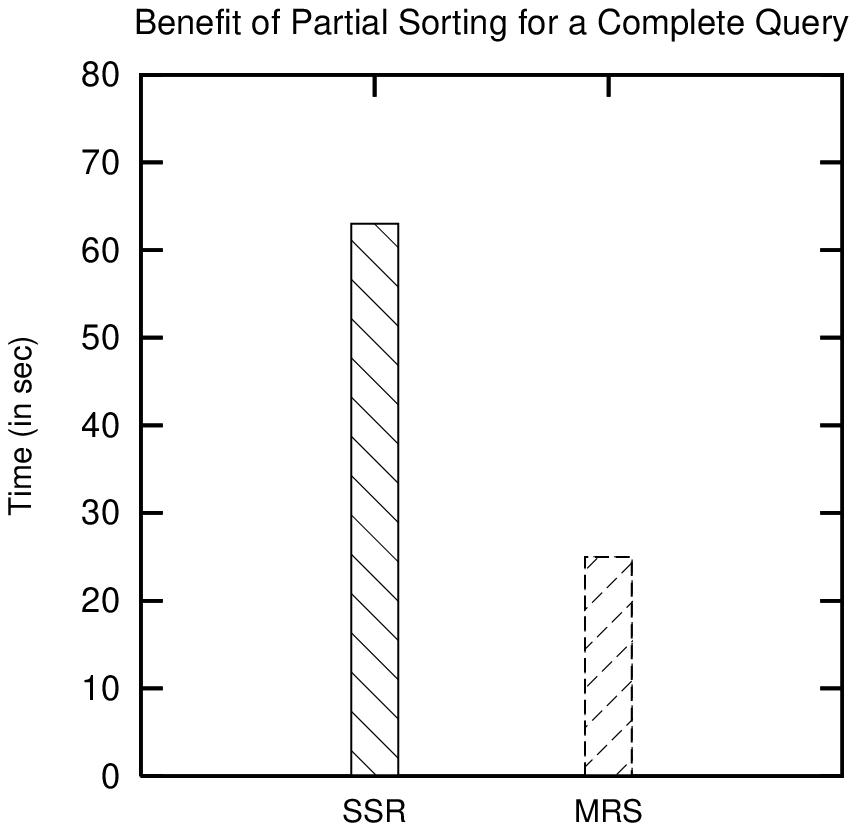}
\caption{Join of partsupp and lineitem} 
\label{fig:perf-pslijoin}
\end{figure}
}

\SubSection{Choice of Interesting Orders}

We extended our Volcano-style cost based optimizer, which we call PYRO,
to consider partial 
sort orders and choose good interesting sort orders for merge joins and 
aggregation. We compare the plans produced by the extended implementation,
which we call PYRO-O, with those of Postgres, SYS1 and SYS2. 
%The optimization overheads due to the extensions were negligible. 

\noindent {\bf Experiment B1}: For this experiment we used Query~\ref{query:outofstock} 
given below, which
lists parts for which the outstanding order quantity is more than the stock
available at the supplier.

\paper{
\begin{query}
\label{query:outofstock}
Parts Running Out of Stock
{\footnotesize
\begin{tabbing}
xxxxxxxxx\=\kill
SELECT \> ps$\_$suppkey, ps$\_$partkey, ps$\_$availqty, sum(l$\_$quantity) \\
%AS total$\_$required \\
FROM  \>  partsupp, lineitem  \\
WHERE \> ps$\_$suppkey=l$\_$suppkey AND ps$\_$partkey=l$\_$partkey AND \\
       \> l$\_$linestatus='O' \\
GROUP  BY ps$\_$availqty, ps$\_$partkey, ps$\_$suppkey \\
HAVING  sum(l$\_$quantity) $>$ ps$\_$availqty ORDER BY ps$\_$partkey;
\end{tabbing}
}
\end{query}
}
\techreport{
\begin{query}
\label{query:outofstock}
Parts Running Out of Stock
{\footnotesize
\begin{tabbing}
xxxxxxxxx\=\kill
SELECT \> ps$\_$suppkey, ps$\_$partkey, ps$\_$availqty, sum(l$\_$quantity) \\
%AS total$\_$required \\
FROM  \>  partsupp, lineitem  \\
WHERE \> ps$\_$suppkey=l$\_$suppkey AND ps$\_$partkey=l$\_$partkey AND l$\_$linestatus='O' \\
GROUP  BY ps$\_$availqty, ps$\_$partkey, ps$\_$suppkey \\
HAVING  sum(l$\_$quantity) $>$ ps$\_$availqty ORDER BY ps$\_$partkey;
\end{tabbing}
}
\end{query}
}

\eat{
\noindent {\bf Query 3:} Parts Running Out of Stock
{\footnotesize
\begin{tabbing}
xxxxxxxxx\=\kill
SELECT \> ps$\_$suppkey, ps$\_$partkey, ps$\_$availqty, sum(l$\_$quantity) \\
%AS total$\_$required \\
FROM  \>  partsupp, lineitem  \\
WHERE \> ps$\_$suppkey=l$\_$suppkey AND ps$\_$partkey=l$\_$partkey AND \\
       \> l$\_$linestatus='O' \\
GROUP  BY ps$\_$partkey, ps$\_$suppkey, ps$\_$availqty \\
HAVING  sum(l$\_$quantity) $>$ ps$\_$availqty ORDER BY ps$\_$partkey;
\end{tabbing}
}
}
Table {\em partsupp} had clustering index on its primary key 
{\em (ps$\_$partkey, ps$\_$suppkey)}.  Two secondary indices, 
one on $ps\_suppkey$ and the other on $l\_suppkey$ were also 
built on the {\em partsupp} and {\em lineitem} tables respectively. 
The two secondary indices covered all attributes needed for the
query. 
\fullpaper{
The experiment shows the need for cost-based choice of 
interesting orders. The choice of interesting orders for the
join and aggregate are not obvious in this case for the following
reasons:
\begin{enumerate}
\item The explicit order by clause favors the choice of of an order
   where {\em partkey} appears first.
\item The clustering index on {\em partsupp} favors the choice of 
   {\em (partkey, suppkey)} as the interesting order.
\item The secondary indices favor the choice of 
   {\em (suppkey, partkey)} that can be obtained by using a low
   cost partial sort.  Note that this option can be much cheaper 
   due to the size of the {\em lineitem} relation.
\end{enumerate}

Therefore, the optimizer must make a cost-based decision on the 
sort order to use. 
}
\fullpaper{
Figures \ref{fig:perf-query1fig1} and 
\ref{fig:perf-query1fig2} show the plans chosen by Postgres,
PYRO-O, SYS1 and SYS2. 
}
\shortpaper{
Figures \ref{fig:perf-query1fig1} shows the plans chosen by Postgres and
PYRO-O. SYS1 chose a hash-join plan by default. When a merge-join plan
was forced with an optimizer hint, SYS1 selected a plan similar to that
of Postgres, except that it avoided the sort of {\em partsupp} by
using the clustering index. The default plan on SYS2 was same as the 
merge-join plan of SYS1.
}

All plans except the hash-join plan of 
SYS1 and the plan produced by PYRO-O use an expensive 
full sort of 6 million lineitem index entries on 
{\em (partkey, suppkey)}. Further, Postgres uses a hash aggregate
where a sort-based aggregate would have been much
cheaper as the required sort order was available from the 
output of merge-join (note that the functional dependency
$\{$ps$\_$partkey, ps$\_$suppkey$\}\rightarrow\{$ps$\_$availqty$\}$
holds).
\fullpaper{
On SYS1, it was possible to force the use of a merge-join
instead of hash-join and the plan chosen is shown in 
\reffig{fig:perf-query1fig2}(b).
}

\paper{
\begin{figure} [t]
\centerline{\psfig{file=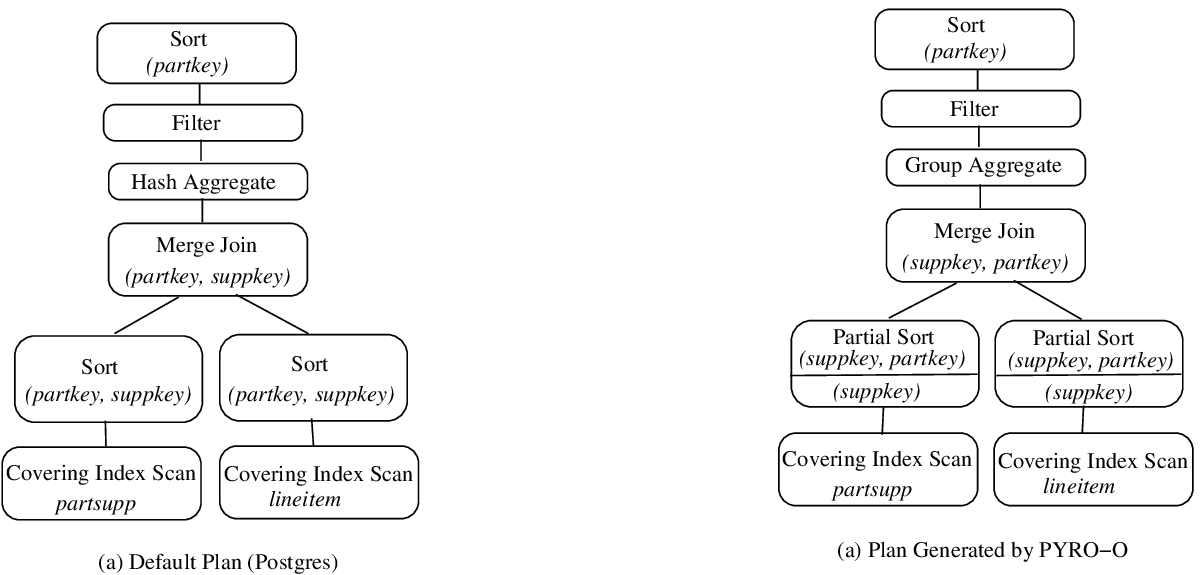}}
\caption{Plans for Query 3 (Postgres and PYRO-O)} 
\label{fig:perf-query1fig1}
\end{figure}
}

\techreport{
\begin{figure} [ht]
%\centerline{\psfig{file=perf/plans/query1fig1.eps}}
\centerline{\includegraphics[angle=0,width=5.0in]{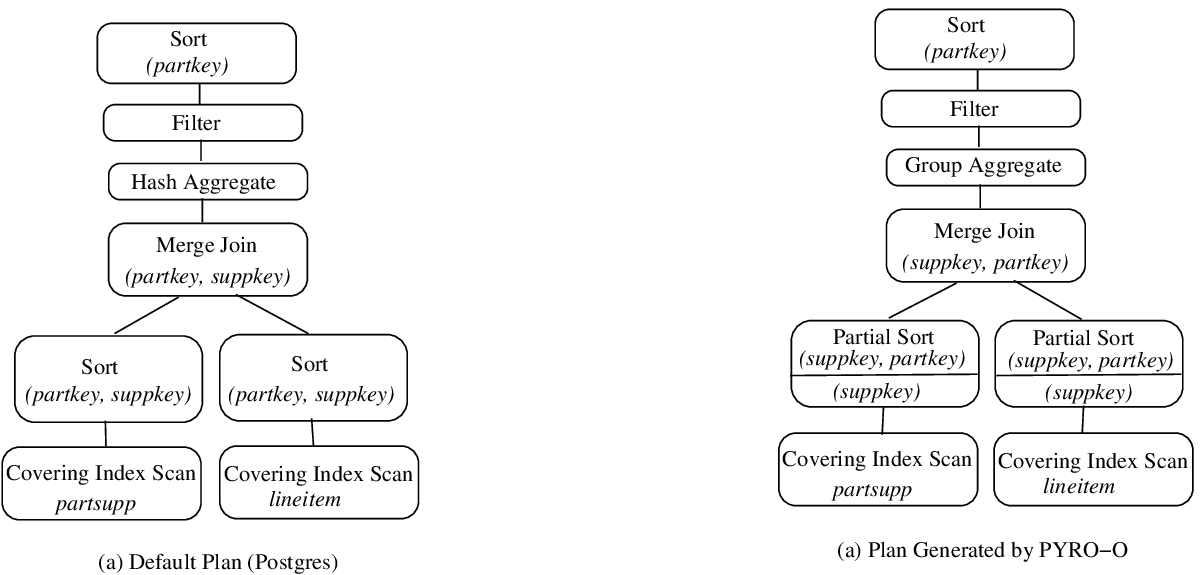}}
\caption{Plans for Query 3 (Postgres and PYRO-O)} 
\label{fig:perf-query1fig1}
\end{figure}
}

\paper{
\fullpaper{
\begin{figure} [t]
\centerline{\psfig{file=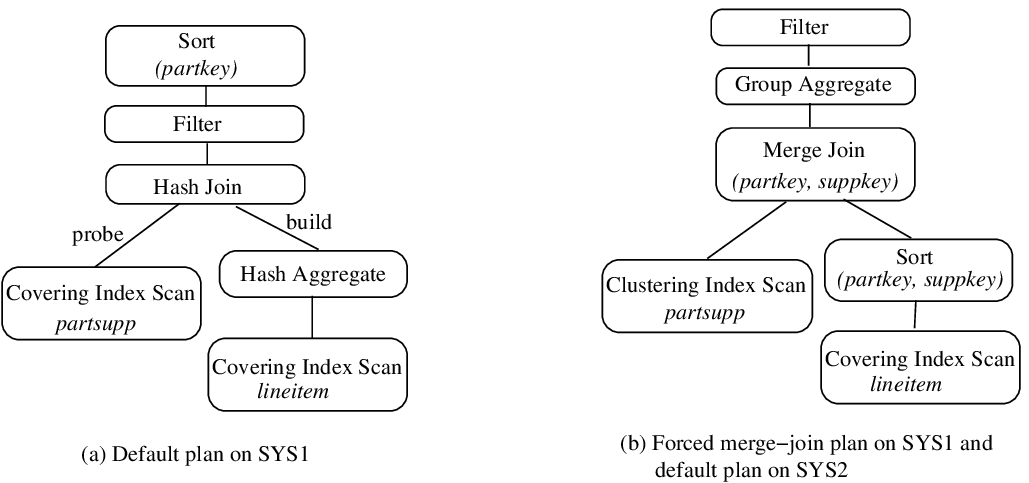}}
\caption{Plans for Query 3 (SYS1 and SYS2)} 
\label{fig:perf-query1fig2}
\end{figure}
}
}
\techreport{
\begin{figure} [t]
%\centerline{\psfig{file=perf/plans/query1fig2.eps}}
\centerline{\includegraphics[angle=0,width=5.0in]{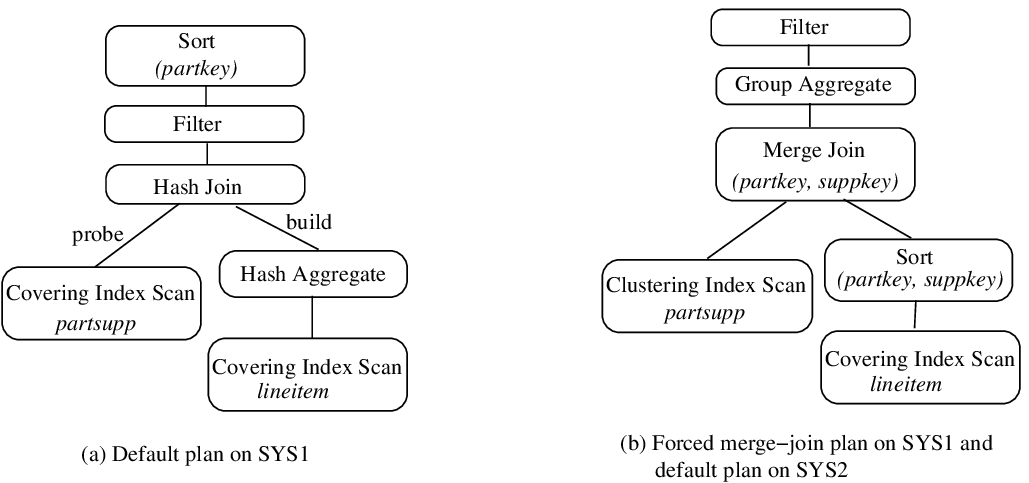}}
\caption{Plans for Query 3 (SYS1 and SYS2)} 
\label{fig:perf-query1fig2}
\end{figure}
}

We compared the actual running time of PYRO-O's plan with those of
Postgres and SYS1 by forcing our plan on the respective systems.
Figures \ref{fig:q2q3postgres} and \ref{fig:q2q3mssql} show
the details.
It was not possible for us to force our plan on SYS2 and make
a fair comparison and hence we omit the same. 
The only surprising
result was the default plan chosen by SYS1 performed slightly
poorer than the forced merge-join plan on SYS1. In all cases,
the forced PYRO-O plan performed significantly better than the
other plans. The main reason for the improvement was the use
of a partial sort of {\em lineitem} index entries as against 
a full
sort. The final sort on {\em partkey} was not very expensive 
as only a few tuples needed to be sorted.

For Query~\ref{query:outofstock} the plan generation phase 
(phase-1) was sufficient to select the sort orders and phase-2
does not make any changes. We shall now see a case for which
phase-1 cannot make a good choice and the sort orders get
refined by phase-2.

\paper{
\begin{figure}[t]
\centering
\begin{tabular}{@{}c@{}c}
\begin{minipage}{0.25\textwidth} %DEFINES THE WIDTH OF EACH PICTURE IN % OF PAGE WIDTH
\includegraphics[angle=270,width=2.4in]{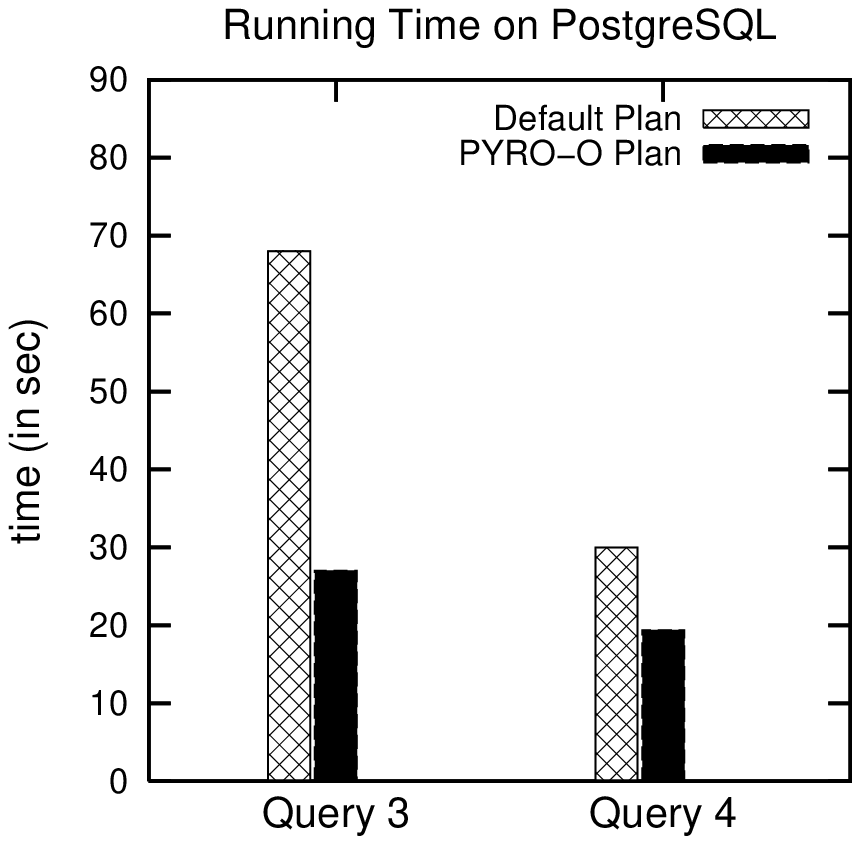}
\caption{PostgreSQL}
\label{fig:q2q3postgres}
\end{minipage}
&
\begin{minipage}{0.25\textwidth}
\includegraphics[angle=270,width=2.4in]{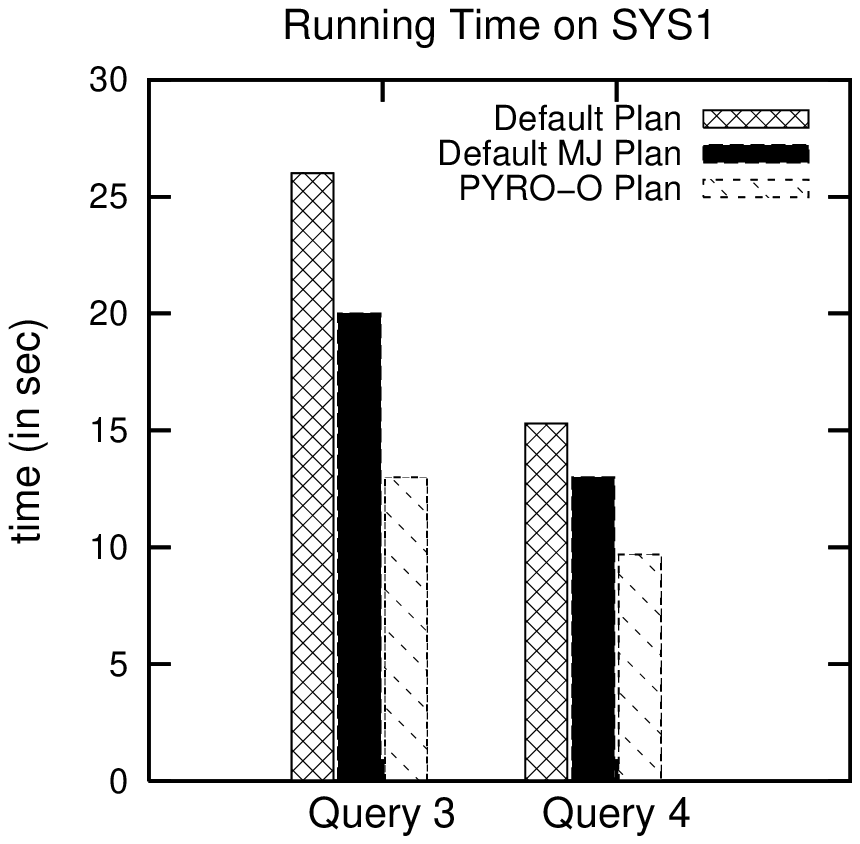}
\caption{SYS1}
\label{fig:q2q3mssql}
\end{minipage}
\end{tabular}
\end{figure}
}

\techreport{
\begin{figure}[ht]
\centering
\begin{tabular}{@{}c@{}c}
\begin{minipage}{0.5\textwidth} %DEFINES THE WIDTH OF EACH PICTURE IN % OF PAGE WIDTH
\includegraphics[angle=270,width=3.4in]{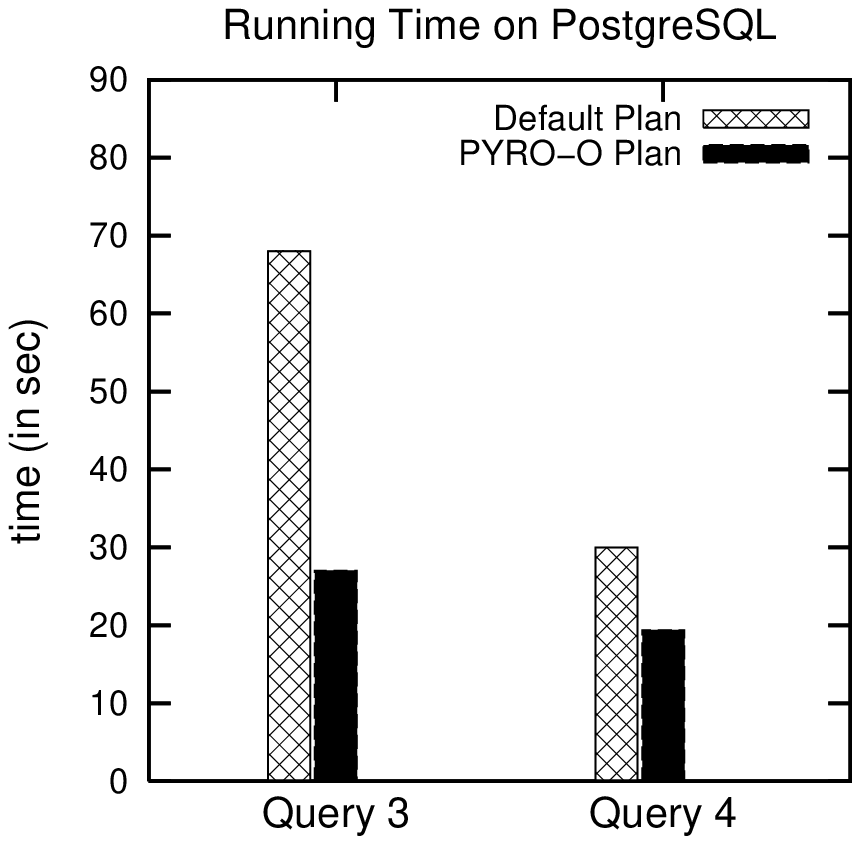}
\caption{PostgreSQL}
\label{fig:q2q3postgres}
\end{minipage}
&
\begin{minipage}{0.5\textwidth}
\includegraphics[angle=270,width=3.4in]{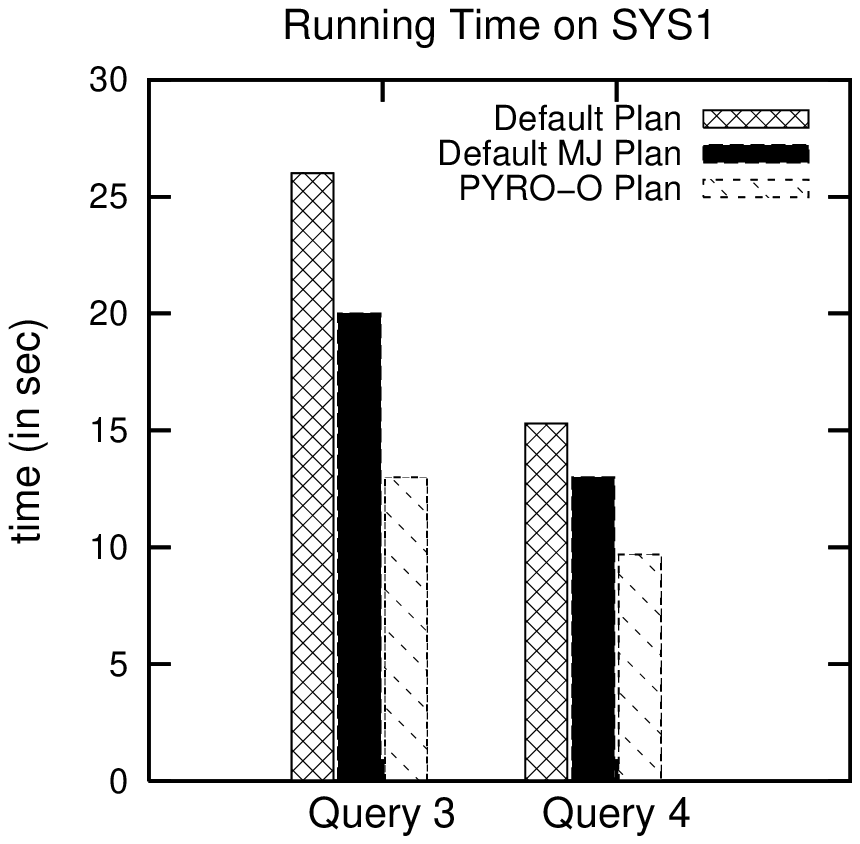}
\caption{SYS1}
\label{fig:q2q3mssql}
\end{minipage}
\end{tabular}
\end{figure}
}

\noindent {\bf Experiment B2}: This experiment uses Query \ref{query:r1-r3},
shown below, which has two full outer
joins with two common attributes between the joins. 
\fullpaper{
We designed this experiment
to see whether the systems we compare with exploit attributes common to multiple
sort-based operators. 
}

\begin{query}
\label{query:r1-r3}
Attributes common to multiple joins
{\footnotesize
\begin{tabbing}
xxxx\=xxxxx\=\kill
SELECT * FROM R1 FULL OUTER JOIN R2 \\
\> ON (R1.c5=R2.c5 AND R1.c4=R2.c4 AND R1.c3=R2.c3) \\
\> FULL OUTER JOIN R3 \\
\> ON (R3.c1=R1.c1 AND R3.c4=R1.c4 AND R3.c5=R1.c5);
\end{tabbing}
}
\end{query}

The tables R1, R2 and R3 were identical and each populated with
100,000 records. No indexes were built. As shown in 
\reffig{fig:perf-r1-r3-sys1-pg-pyro}(a), both SYS1 and Postgres chose
sort orders that do not share any common prefix. The plan chosen 
by PYRO-O is shown in \reffig{fig:perf-r1-r3-sys1-pg-pyro}(b).
In the plan chosen by PYRO-O, the two joins share a common prefix of 
{\em (c4, c5)} and thus the sorting effort is expected to be 
significantly less. 
%\shortpaper{
SYS2, not having an implementation
of full outer join, chose a union of two left outer joins.
The two left outer joins used to get a full outer
join used different sort orders making the union expensive, 
illustrating a need for coordinated choice of sort orders. 
%}

\eat{
\fullpaper{
The plan chosen by SYS2 is shown in 
\reffig{fig:perf-r1-r3-sys2}. SYS2, not having an implementation
of full outer join, choose a union of two left outer joins.
The two left outer joins between R1 and R2 use 
different sort orders. This, not only takes away the opportunity
of reusing intermediate results (sorted R1 and R2) but also 
makes the the UNION operator more expensive. Note that a
one-pass merge UNION requires the same sort order from both
inputs. 
}
}

\paper{
\begin{figure}
%\centerline{\psfig{file=perf/plans/perf-r1-r3-mssql-pg-pyro.eps}}
\begin{center}
\includegraphics[angle=0,height=1.6in,width=2.8in]{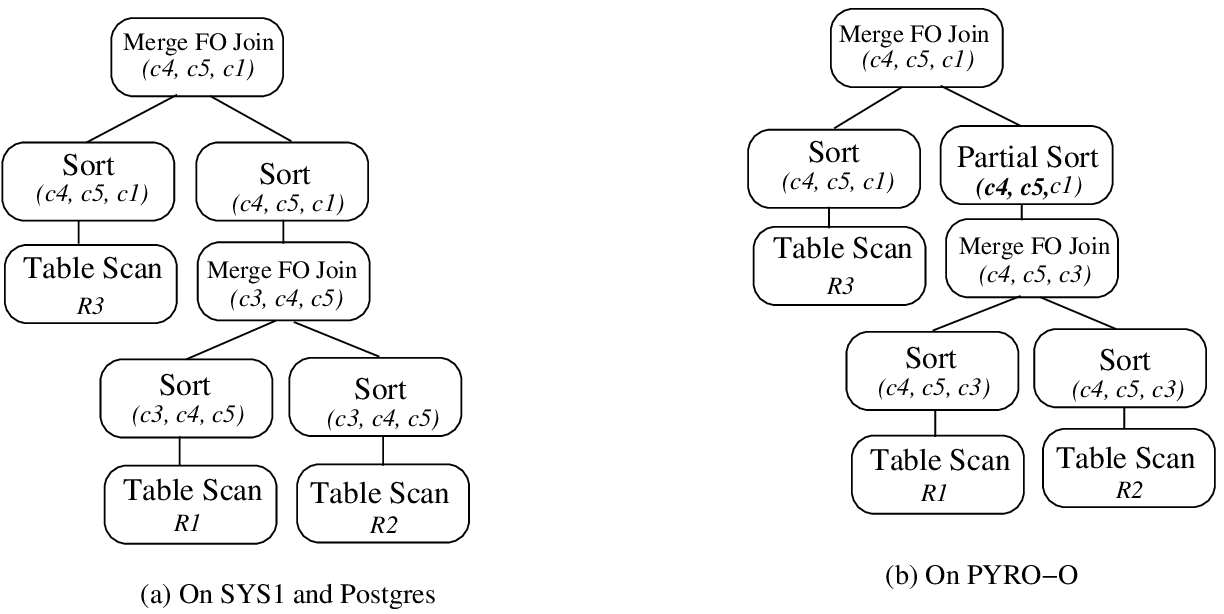}
\end{center}
\vspace{-2ex}
\caption{Plans for Query 4} 
\label{fig:perf-r1-r3-sys1-pg-pyro}
\end{figure}

%\fullpaper{
%\begin{figure} [t]
%\centerline{\psfig{file=perf/plans/perf-r1-r3-db2.eps}}
%\caption{Plan for Query 4 on SYS2} 
%\label{fig:perf-r1-r3-sys2}
%\end{figure}
%}
}

\techreport{
\begin{figure}
%\centerline{\psfig{file=perf/plans/perf-r1-r3-mssql-pg-pyro.eps}}
\begin{center}
\includegraphics[angle=0,height=2.4in,width=5.5in]{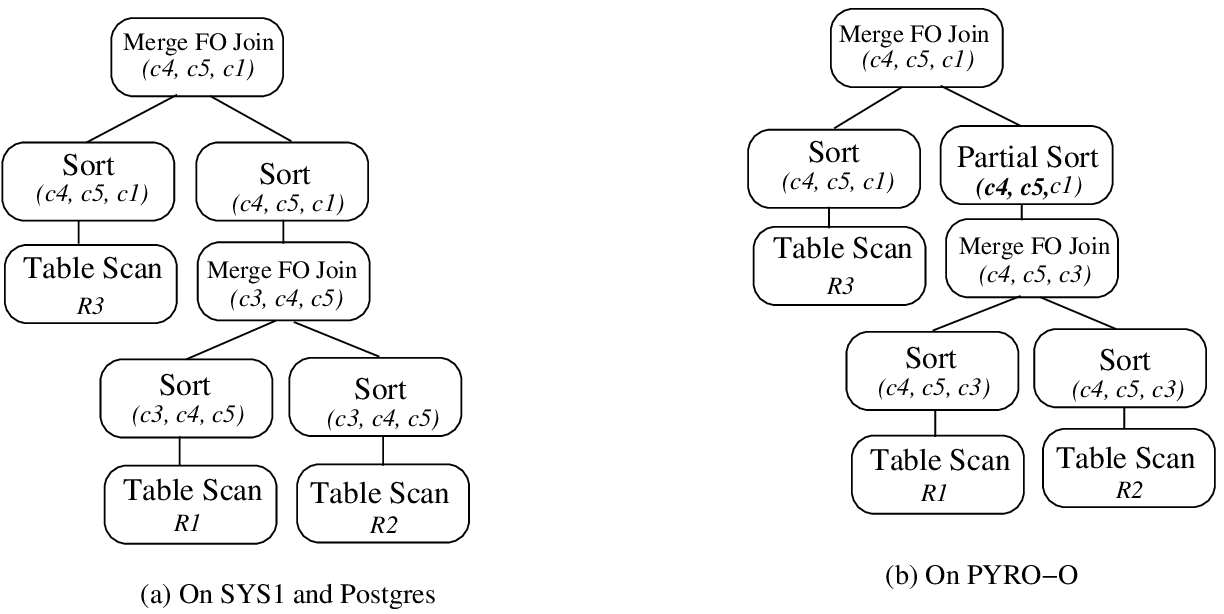}
\end{center}
\vspace{-2ex}
\caption{Plans for Query 4} 
\label{fig:perf-r1-r3-sys1-pg-pyro}
\end{figure}

%\fullpaper{
%\begin{figure} [t]
%\centerline{\psfig{file=perf/plans/perf-r1-r3-db2.eps}}
%\caption{Plan for Query 4 on SYS2} 
%\label{fig:perf-r1-r3-sys2}
%\end{figure}
%}
}

\noindent {\bf Experiment B3}: In this experiment we compare our approach
of choosing interesting orders, PYRO-O, with the exhaustive approach,
 and a heuristic used by PostgreSQL.
Postgres uses the following heuristic: for each of the $n$ attributes
involved in the join condition, a sort order beginning with that attribute 
is chosen; in each order, the remaining $n-1$ attributes are ordered
arbitrarily. We implemented Postgres' heuristic in PYRO along with the
extensions to exploit partial sort orders and call it PYRO-P. The 
exhaustive approach, called PYRO-E, enumerates all $n!$ permutations
and considers partial sort orders. In addition, we also compare with
PYRO, which chooses an arbitrary sort order, 
and a variation of PYRO-O, called PYRO-O$^-$ that considers only exact 
favorable orders (no partial sort).
\reffig{fig:perf-plancost} shows the estimated plan costs. Note the
logscale for y-axis. The plan
costs are normalized taking the plan cost with exhaustive approach
to be 100. In the figure, Q3 and Q4 are Query~\ref{query:outofstock}
and Query \ref{query:r1-r3} of Experiments B1 and B2. 
\shortpaper{Q5 and Q6 were
two real-world analytical queries and can be found in the full
length paper~\cite{FULLPAPER}.
}
\fullpaper{
Q5 and Q6 are Queries~\ref{query:ibank-trans} and ~\ref{query:ibank-basket}
shown below.
} 
For Q3 and Q4, as very few attributes
were involved in the join condition, Postgres' heuristic
along with extensions to exploit partial sort orders, produced
plans which were close to optimal. However, for more complex
queries the heuristic does not perform as well since it makes
an arbitrary choice for secondary orders. 

\paper{
\fullpaper{
\begin{query}
\label{query:ibank-trans}
Total value executed for a given order
{\footnotesize
\begin{tabbing}
xxxxx\=xxxxx\=\kill
SELECT T1.UserId, T1.BasketId, T1.ParentOrderId, T1.WaveId, \\
    \> T1.ChildOrderId, (T1.Quantity * T1.Price) as OrderValue, \\
    \> SUM(T2.Quantity * T2.Price) as ExecutedValue \\
FROM TRAN T1, TRAN T2 \\
WHERE T1.UserId=T2.UserId AND T1.ParentOrderId=T2.ParentOrderId \\
    \> AND T1.BasketId=T2.BasketId AND T1.WaveId=T2.WaveId \\
    \> AND T1.ChildOrderId=T2.ChildOrderId AND T1.TranType='New' \\
    \> AND T2.TranType='Executed' \\
GROUP BY T1.UserId, T1.BasketId, T1.ParentOrderId, T1.WaveId, \\
    \> T1.ChildOrderId;
\end{tabbing}
}
\end{query}

\begin{query}
\label{query:ibank-basket}
Basket Analytics
{\footnotesize
\begin{tabbing}
xxxxx\=xxxxx\=\kill
SELECT * FROM BASKET B, ANALYTICS A \\
WHERE B.PRODTYPE = A.PRODTYPE AND B.SYMBOL = A.SYMBOL \\
\> AND B.EXCHANGE = A.EXCHANGE
\end{tabbing}
}
\end{query}
}
}

\techreport{
\fullpaper{
\begin{query}
\label{query:ibank-trans}
Total value executed for a given order
{\footnotesize
\begin{tabbing}
xxxxxxxx\=xxxxx\=\kill
SELECT T1.UserId, T1.BasketId, T1.ParentOrderId, T1.WaveId, T1.ChildOrderId, \\
    \> (T1.Quantity * T1.Price) as OrderValue, SUM(T2.Quantity * T2.Price) as ExecutedValue \\
FROM TRAN T1, TRAN T2 \\
WHERE T1.UserId=T2.UserId AND T1.ParentOrderId=T2.ParentOrderId AND T1.BasketId=T2.BasketId \\
    \> AND T1.WaveId=T2.WaveId AND T1.ChildOrderId=T2.ChildOrderId AND T1.TranType='New' \\
    \> AND T2.TranType='Executed' \\
GROUP BY T1.UserId, T1.BasketId, T1.ParentOrderId, T1.WaveId, T1.ChildOrderId;
\end{tabbing}
}
\end{query}

\begin{query}
\label{query:ibank-basket}
Basket Analytics
{\footnotesize
\begin{tabbing}
xxxxx\=xxxxx\=\kill
SELECT * FROM BASKET B, ANALYTICS A \\
WHERE B.ProdType = A.ProdType AND B.Symbol = A.Symbol 
AND B.Exchange = A.Exchange;
\end{tabbing}
}
\end{query}
}
}

\paper{
\begin{figure}[t]
%\centering
\hspace{-3ex}
\begin{tabular}{@{}c@{}c}
\begin{minipage}{0.25\textwidth} %DEFINES THE WIDTH OF EACH PICTURE IN % OF PAGE WIDTH
\includegraphics[angle=270,totalheight=1.5in,width=2.2in]{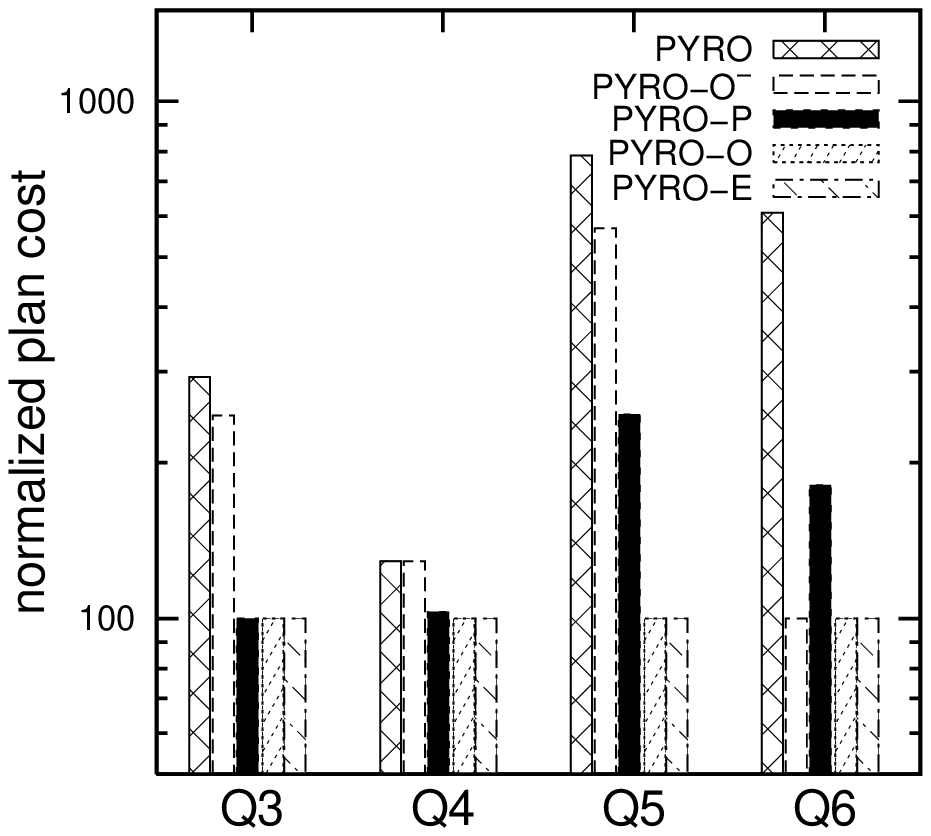}
%\caption{Gains of Partial Sorting} 
\caption{More Queries} 
\label{fig:perf-plancost}
\vspace{-10pt}
\end{minipage}
&
\begin{minipage}{0.25\textwidth} %DEFINES THE WIDTH OF EACH PICTURE IN % OF PAGE WIDTH
\includegraphics[angle=270,totalheight=1.6in,width=2.2in]{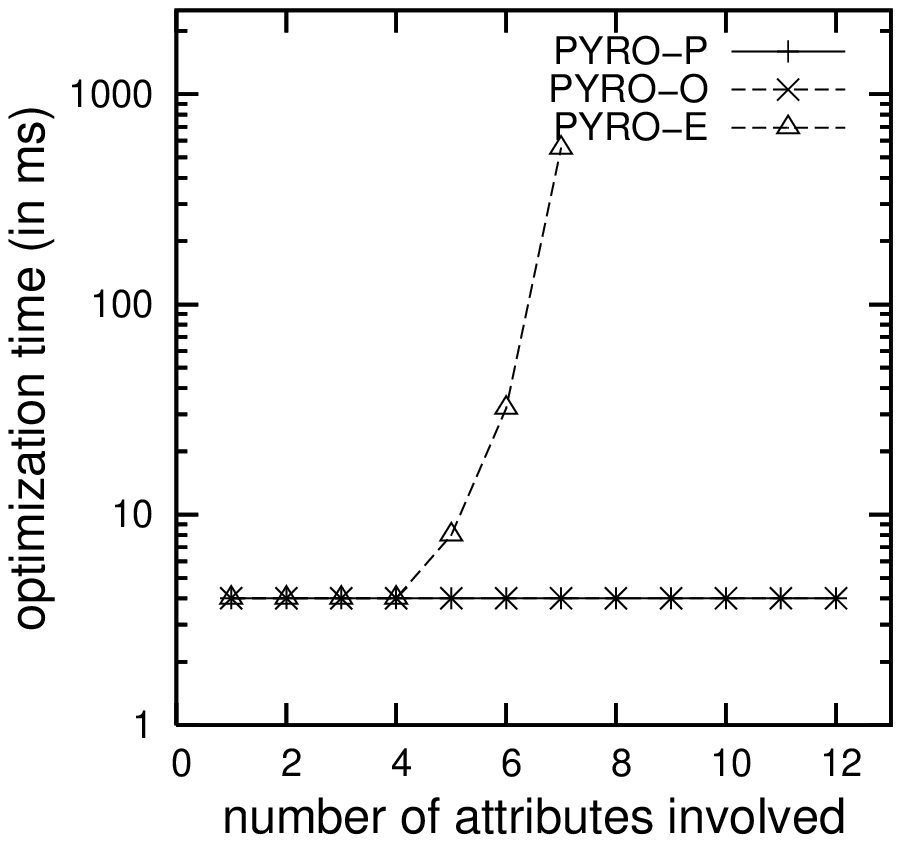}
\caption{Scalability}
\label{fig:perf-opttime}
\end{minipage}
\end{tabular}
\end{figure}
}

\techreport {
\begin{figure}[ht]
%\centering
\hspace{-3ex}
\begin{tabular}{@{}c@{}c}
\begin{minipage}{0.5\textwidth} %DEFINES THE WIDTH OF EACH PICTURE IN % OF PAGE WIDTH
\centerline {\includegraphics[angle=270,totalheight=2.0in,width=3.4in]{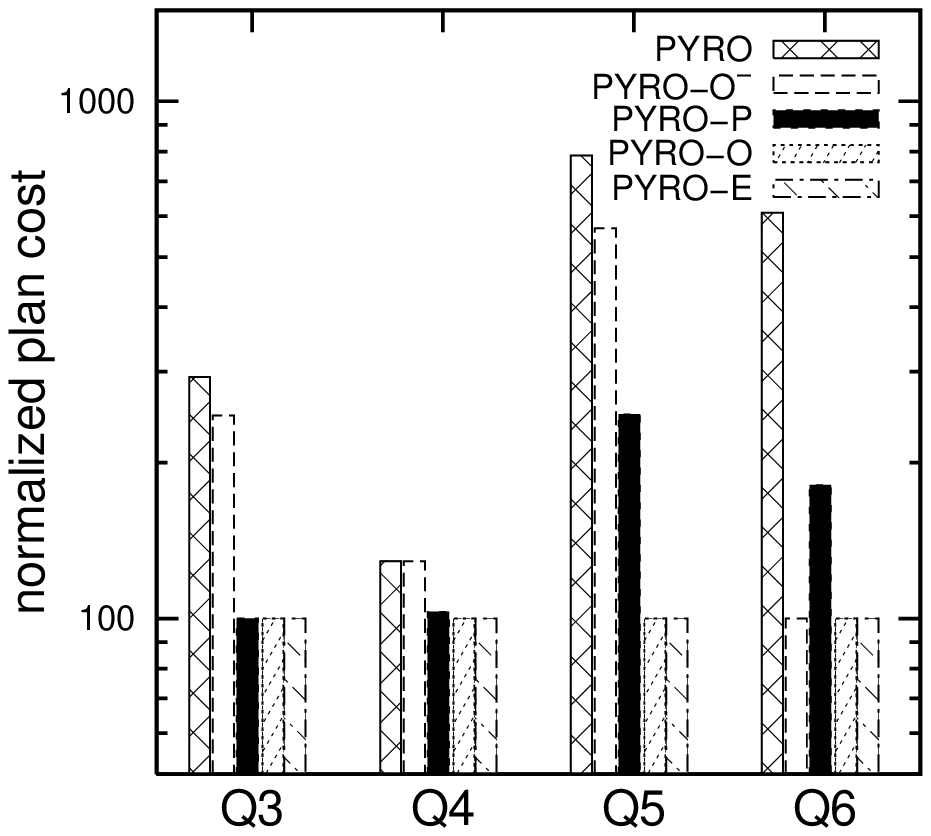}}
%\caption{Gains of Partial Sorting} 
\caption{More Queries} 
\label{fig:perf-plancost}
\vspace{-10pt}
\end{minipage}
&
\begin{minipage}{0.5\textwidth} %DEFINES THE WIDTH OF EACH PICTURE IN % OF PAGE WIDTH
\centerline{\includegraphics[angle=270,totalheight=2.0in,width=3.4in]{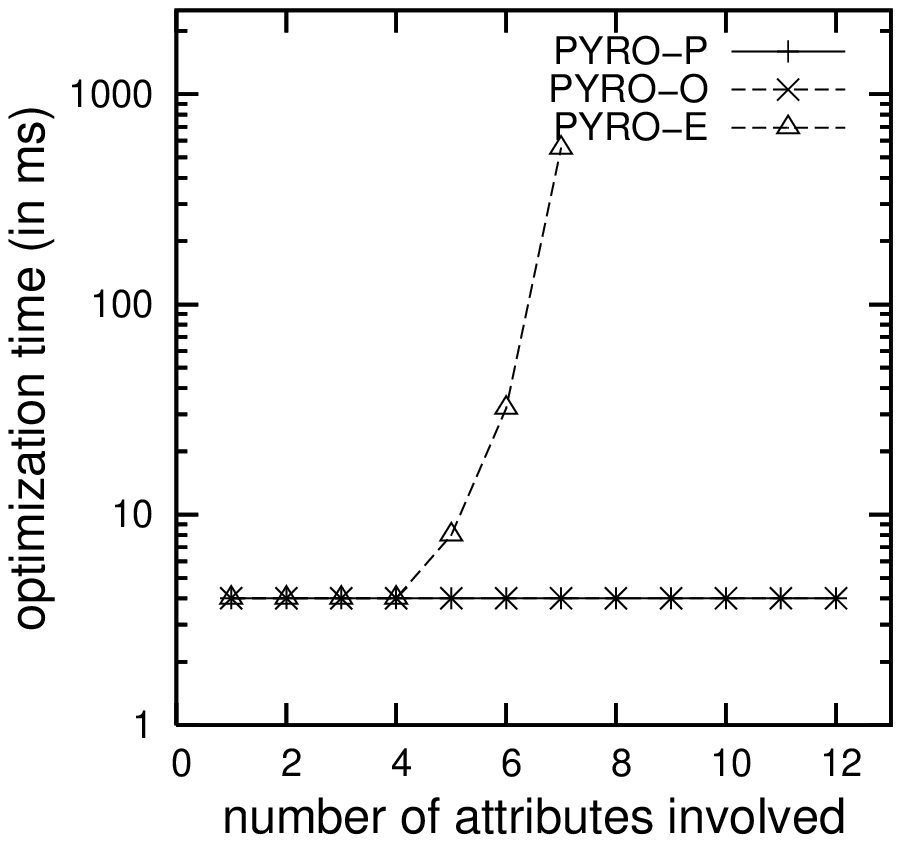}}
\caption{Scalability}
\label{fig:perf-opttime}
\end{minipage}
\end{tabular}
\end{figure}
}

\SubSection {Optimization Overheads}
% and Implementation Complexity}
The optimization overheads due to the proposed extensions were negligible. 
During plan generation, the number of interesting orders we try at 
each join or aggregate node is of the order of the number of indices
that are useful for answering the query. We found that in most real-life
queries this number is fairly small.
% (less than 4). 
\reffig{fig:perf-opttime} shows the scalability of the three heuristics.
For the experiment a query that joined two relations on varying number
of attributes was used. Though PYRO-P and PYRO-O take the same amount
of time in this experiment, in most cases, the number of favorable
orders is much less than the total number of attributes involved
and hence PYRO-O generates significantly fewer interesting orders than 
PYRO-P. 

The plan-refinement algorithm presented in \refsec{subsec:approxalgo} 
was tested with trees up to 31 nodes (joins) and 10 attributes per node. The 
time taken was negligible in each case. The execution of plan 
refinement phase took less than 6 ms even for the tree with 31 nodes.

Both the optimizer extensions and the extension to external-sorting (MRS) 
were fairly straight forward to implement. 
%Our changes to PostgreSQL's sort were completed in a week's time. 
The optimizer extensions neatly integrated into our existing Volcano style 
optimizer.
%though involving more code, 

%;. In our approach, the sort order requirements on subexpressions
%are concrete and we do not carry around a flexible order requirement. 
%This reduces the implementation complexity significantly.

%\input{section2.tex}
\paper{
\Section{Conclusion} 
}
\techreport{
\Section{Conclusion and Future Work}
}
\label{sec:concl}
We addressed the issue of choosing interesting 
sort orders.  We showed that even a simplified version of the problem
is {\em NP-hard} and proposed principled heuristics for choosing interesting orders. 
Our techniques take into account important issues 
such as partially matching sort orders and operators that 
require matching sort orders from multiple inputs. We presented detailed
experimental results to demonstrate the benefits due to our techniques.

\paper{
Unlike merge-join and order-by, operators such as group-by and 
duplicate elimination actually need grouped but not necessarily
sorted input; sorting is 
just one way of providing grouped input. Extending our techniques
to grouped input property is a topic of future work. 
}

\techreport{
%\subsection*{Future Work}
Secondary indices that do not cover the query can also be efficient
to obtain a desired sort order if the actual tuple fetch can be 
deferred. Deferring the fetch until a point where the extra attributes
are actually needed can be very effective when a highly selective
filter discards many rows before the fetch is needed. Such optimizations
must be made in a cost-based manner.

Next, we would like to investigate additional ways of speeding up 
nested iteration involving user-defined functions. Though decorrelation
is not always applicable for user-defined functions, we expect many
cases that can be unnested using a combination of dataflow analysis 
and known unnesting methods.

In many cases users do not retrieve the whole of the query result but
need only the first few tuples. In Top-K queries this interest is explicitly
stated and the query optimizers attempt to find a plan that is optimal 
for returning the first K results. However, there are cases where the
value of K is not known a priori. The user may decide to terminate the
query execution at an arbitrary point, perhaps based on the results
seen so far. We would like to study plans that adapt with the number 
of rows retrieved.
}

\nocite{ACM:QEVAL}
\nocite{KNUTH}
\nocite{VLDB05}

%------------------------------------------------------------------------- 
\paper{
\bibliographystyle{latex8}
}
\techreport{
\bibliographystyle{alpha}
}
\bibliography{eiotechrep}

\newcommand{\etalchar}[1]{$^{#1}$}
\begin{thebibliography}{SMD{\etalchar{+}}79}
\setlength{\itemsep}{-0.25ex}

\bibitem[DPS02]{ACM:LAYOUTSUR}
Josep Diaz, Jordi Petit, and Maria Serna.
\newblock {A} {S}urvey of {G}raph {L}ayout {P}roblems.
\newblock {\em ACM Comput. Surv.}, 34(3), 2002.

\bibitem[ECW92]{WOOD}
Vladimir Estivill-Castro and Derick Wood.
\newblock A survey of adaptive sorting algorithms.
\newblock {\em ACM Comput. Surv.}, 24(4), 1992.

\bibitem[GM93]{GRA:ICDE93}
Goetz Graefe and W.J. McKenna.
\newblock {T}he {V}olcano {O}ptimizer {G}enerator: {E}xtensibility and
  {E}fficient {S}earch.
\newblock In {\em ICDE}, 1993.

\bibitem[Gra93]{ACM:QEVAL}
Goetz Graefe.
\newblock {Q}uery {E}valuation {T}echniques for {L}arge {D}atabases.
\newblock {\em ACM Comput. Surv.}, 25(2), 1993.

\bibitem[GRS05]{VLDB05}
Ravindra Guravannavar, H.~S. Ramanujam, and S.~Sudarshan.
\newblock {O}ptimizing {N}ested {Q}ueries with {P}arameter {S}ort {O}rders.
\newblock In {\em VLDB}, pages 481--492, 2005.

\bibitem[Knu73]{KNUTH}
D.~Knuth.
\newblock {\em The Art of Programming, Vol. 3 (Sorting and Searching)}.
\newblock Addison-Wesley, 1973.

\bibitem[Lar03]{PAUL}
Per-{\AA}ke Larson.
\newblock External sorting: Run formation revisited.
\newblock {\em IEEE Trans. Knowl. Data Eng.}, 15(4), 2003.

\bibitem[NM04a]{MOER:VLDB04}
Thomas Neumann and Guido Moerkotte.
\newblock {A} {C}ombined {F}ramework for {G}rouping and {O}rder {O}ptimization.
\newblock In {\em VLDB}, 2004.

\bibitem[NM04b]{MOER:ICDE04}
Thomas Neumann and Guido Moerkotte.
\newblock {A}n {E}fficient {F}ramework for {O}rder {O}ptimization.
\newblock In {\em ICDE}, 2004.

\bibitem[SMD{\etalchar{+}}79]{SEL:SIGMOD79}
P.~Griffiths Selinger, M.M.Astrahan, D.D.Chamberlin, R.A.Lorie, and T.G.Price.
\newblock {A}ccess {P}ath {S}election in a {R}elational {D}atabase {M}anagement
  {S}ystem.
\newblock In {\em Proceedings of ACM SIGMOD Conference}, 1979.

\bibitem[SSM96]{MALK:SIGMOD96}
David Simmen, Eugene Shekita, and Timothy Malkemus.
\newblock {F}undamental {T}echniques for {O}rder {O}ptimization.
\newblock In {\em Proceedings of ACM SIGMOD Conference}, 1996.

\bibitem[WC03]{MITCH:VLDB03}
Xiaoyu Wang and Mitch Cherniack.
\newblock {A}voiding {S}orting and {G}rouping {I}n {P}rocessing {Q}ueries.
\newblock In {\em VLDB}, 2003.

\end{thebibliography}

\fullpaper{
\appendix
\section{Optimality with Exact Favorable Orderes} 
\label{sec:proof}
\begin{theorem}
\label{claim1}
The set $\mathcal{I}(e,o)$ computed with exact ford-mins contains an 
optimal sort order $o_p$ for the 
%goal ${\mathcal G}$ to 
optimize goal $e=(e_l \Join e_r)$ with $(o)$ as the required output 
sort order.

\noindent Assumption: \\
If $o_1$, $o_2$ are two orders such that attrs$(o_1)=$attrs$(o_2)$, 
cpu-cost$(e, o_1)=$cpu-cost$(e, o_2)$
%~\symbolfootnote[1]{Though this is not a very accurate cost model, we restrict ourselves to this model 
%as the problem is hard even for this simple model.} 
\end{theorem}

\begin{proof}
We define the plan cost of an order $o'$ for the optimization goal 
$(e=e_l\Join e_r, o)$, where $o$ is the required output order as follows:

\noindent
%{\small
{\boldmath $PC(e, o, o')$}$=cbp(e_l, o') + cbp(e_r, o') + coe(e, o', o) + CM$,
%} 
where CM is the cost of the merge join operator, which we assume to be same for 
all $o'$ such that $attrs(o')=S$, where $S$ is the set of attributes involved in 
the join predicate.

\noindent
Let $o_b$ be an {\em optimal order} not belonging to set $\mathcal{I}(e)$. We show 
that $\exists o_p\in \mathcal{I}(e)$ such that {\em PC}$(o_p)=${\em PC}$(o_b)$

\vspace{0.2in}
\noindent {\bf Case 1:} $o_b \notin ford(e_l) \bigcup ford(e_r)$ \\
Let $o_p$ be an order $\in \mathcal{I}(e)$ such that   $o \wedge S \leq o_p$. Note that there must
exist such an order in $\mathcal{I}(e)$ since $o \wedge S \in \mathcal{T}$ (Step 1 in computing $\mathcal{I}(e)$) \\
\vspace{-0.1in}
%{\small
\begin{tabbing}
xxxxxxx\=\kill
$PC(o_b)$ \> $= cbp(e_l, o_b) + cbp(e_r, o_b) + coe(e, o_b, o) + CM$ \\
Since, $o_b \notin ford(e_l) \bigcup ford(e_r)$ we can write \\
$PC(o_b)$ \> $= cbp(e_l, \epsilon) + coe(e_l, \epsilon, o_b) + cbp(e_r, \epsilon) +$ \\
             \> $coe(e_r, \epsilon, o_b) + coe(e, o_b, o) + CM$ \\
%$cbp(e_l, o_b) = cbp(e_l, \epsilon) + coe(e_l, \epsilon, o_b)$ and 
%$cbp(e_r, o_b) = cbp(e_r, \epsilon) + coe(e_r, \epsilon, o_b)$ \\
%\noindent Under Tuple Comparison Model-1, $coe(e_x, \epsilon, o_b)=coe(e_x, \epsilon, o_p)$ 
%since $||o_b||=||o_p||$. Therefore, \\
Since {\em attrs}$(o_b)$={\em attrs}$(o_p)$, $coe(e_x, \epsilon, o_b)=coe(e_x, \epsilon, o_p)$. Hence,  \\
$PC(o_b)$ \> $= cbp(e_l, \epsilon) + coe(e_l, \epsilon, o_p) + cbp(e_r, \epsilon) + $\\
             \> $\ \ \ \ coe(e_r, \epsilon, o_p) + coe(e, o_b, o) + CM$ \\

            \> $\geq cbp(e_l, o_p) + cbp(e_r, o_p) + coe(e, o_b, o) + CM$ \\
Further, $o_b \wedge o \leq o_p \wedge o$ since $o \wedge S \leq o_p$ and $o_b$ is a permutation\\
of $S$. Hence,\\
%From {\em Equation \ref{eqn:coe1}}, we conclude:\\
$PC(o_b)$ \> $\geq cbp(e_l, o_p) + cbp(e_r, o_p) + coe(e, o_p, o) + CM$ \\
$PC(o_b)$ \> $\geq PC(o_p)$.
\end{tabbing}
%}
\vspace{0.2in}

\noindent {\bf Case 2:} $o_b \in ford(e_l) \bigcup ford(e_r)$ \\

\noindent
{\bf Case 2A:} $o_b \in ford(e_l)$ OR $o_b \in ford(e_r)$ but not both. Without loss 
of generality assume $o_b \in ford(e_l)$. This implies one of the following:

\begin{enumerate}
    \item $\exists o' \in$ {\em ford-min}$(e_l)$ such that $o_b \leq o'$ \\
    $o' \in$ {\em ford-min}$(e_l) \Longrightarrow \exists o_p \in \mathcal{I}(e)$ such that $o' \wedge S \leq o_p$ \\
    Since $o_b$ is a permutation of $S$ we have $o_b \wedge S = o_b$ \\
    Therefore $o' \wedge S = o_b$ and hence $o_b <= o_p$ \\
    Further, since $|o_b| = |o_p|$, we conclude $o_b=o_p$. This contradicts our earlier 
    assumption that $o_b \notin \mathcal{I}(e)$

   \item $\exists o' \in$ {\em ford-min}$(e_l)$ such that $o' \leq o_b$ and $cbp(e_l, o') + 
           coe(e_l, o', o_b) = cbp(e_l, o_b)$ \\
    $o' \in$ {\em ford-min}$(e_l) \Longrightarrow \exists o_p \in \mathcal{I}(e)$ such that  $o' \wedge S \leq o_p$

%{\small
\begin{tabbing}
x\=xxxxxxxx\=xx\=\kill
    \> $PC(e, o_b)$ \> $=  cbp(e_l, o_b) + cbp (e_r, o_b) + coe (e, o_b, o) + CM$ \\
        \>  \>  $= cbp(e_l, o') + coe(e_l, o', o_b) + cbp(e_r, o_p) +$ \\
        \>  \>  $\ \ coe (e, o_b, o) + CM$ \\
        \>  (Note: $cbp(e_r, o_b) = cbp(e_r, o_p)$ as $o_b \notin ford(e_r)$) \\
        \>  Since $attrs(o_b)=attrs(o_p)$ we can write \\
    \> $PC(e, o_b)$ \>  $= cbp(e_l, o') + coe(e_l, o' o_p) +  cbp(e_r, o_p) + $\\
    \>                  \> $\ \ \ \ coe (e, o_b, o) + CM$   \\
        \>  \>  $\geq cbp(e_l, o_p) + cbp(e_r, o_p) + coe(e, o_b, o) + CM$ \\
\\
        \> {\bf case (a)} $o' \leq o$ \\
        \> Then we could choose $o_p$ from $\mathcal{I}(e)$ such that \\
        \> $o \leq o_p$. Hence, $coe(e, o_b, o) \geq coe(e, o_p, o)$ \\
        \> $PC(e, o_b)$  \>  \> $\geq  cbp(e_l, o_p) + cbp(e_r, o_p) + coe(e, o_p, o) + CM$ \\
        \>  \> \>   $\geq PC(e, o_p)$ \\
\\
        \> {\bf case (b)} $o' \nleq o$ \\
        \>  Now, $o' \wedge  o = o_b \wedge  o = o_p \wedge o$ ($\because o' \leq o_b$ and $o' \leq o_p$) \\
        \> $PC(e, o_b)$ \> $\geq cbp(e_l, o_p) + cbp(e_r, o_p) + coe(e, o_p, o) + CM$ \\
        \>  \>  $\geq PC(e, o_p)$ \\
\end{tabbing}
%}
\end{enumerate}

\noindent
{\bf Case 2B:} $o_b \in ford(e_l)$  as well as $o_b \in ford(e_r)$ \\
This implies one of the following:

\begin{enumerate}
    \item $\exists o' \in$ {\em ford-min}$(e_l)$ (or {\em ford-min}$(e_r)$) such that $o_b \leq o'$ \\
    In this case the proof can proceed as in {\bf Case 2A}(1).

    \item $\exists o_1, o_2 \in$ {\em ford-min}$(e_l)$ such that $o_i \leq o_b$ and $cbp(e_l, o_i) + coe(e_l, o_i, o_b) = cbp(e_l, o_b)$ (for both $i=1$ and $i=2$) \\
      Since $o_1 \leq o_b$ and $o_2 \leq o_b$, either $o_1 \leq o_2$ or $o_2 \leq o_1$. Hence,
      $\exists o_p \in \mathcal{I}(e)$ such that $o_1\wedge S \leq o_p$ and $o_2 \wedge S \leq o_p$\\
      Choosing this $o_p$ and proceeding as in {\bf Case 2A} (2) we can prove 
      $PC(e, o_b) \geq PC(e, o_p)$
\end{enumerate}

\end{proof}
}

\end{document}